\documentclass[aps,pre,reprint,twocolumn,superscriptaddress,floatfix]{revtex4-2}

\usepackage{amsmath}
\usepackage{amsfonts}
\usepackage{amssymb}
\usepackage{mathtools}
\usepackage{empheq}
\usepackage{multirow}
\usepackage{indentfirst}
\usepackage{color}

\usepackage{graphicx}
\usepackage[dvipsnames]{xcolor}
\usepackage{subcaption}

\usepackage[varg]{txfonts}
\usepackage{newtxtext}

\usepackage{bm}

\usepackage{physics}

\let\ln\naturallogarithm
\allowdisplaybreaks

\definecolor{pdarkblue}{rgb}{0.1797, 0.1875, 0.5703}
\usepackage{hyperref}
\hypersetup{
    colorlinks=true,
    citecolor=pdarkblue,
    linkcolor=pdarkblue,
    urlcolor=pdarkblue,
}
\usepackage[all]{hypcap}

\AtBeginDocument{
\fontsize{10.5pt}{12.36pt}\selectfont
\fontdimen2\font = 2.9pt 
}
\usepackage{titlesec}
\titleformat*{\section}{\centering\fontsize{10.5pt}{\baselineskip}\selectfont\bfseries}
\titleformat*{\subsection}{\centering\fontsize{10.5pt}{\baselineskip}\selectfont\bfseries}
\titleformat*{\subsubsection}{\centering\fontsize{10.5pt}{\baselineskip}\selectfont\itshape}
\titlespacing*{\section}{\linewidth}{2.5em}{0.4em}
\titlespacing*{\subsection}{\linewidth}{1.5em}{0.4em}
\titlespacing*{\subsubsection}{\linewidth}{1.5em}{0.4em}

\newcommand{\red}[1]{\textcolor{black}{#1}}

\begin{document}

\title{Heat dissipation in marginally stable linear time-delayed Langevin systems}
\date{\today}

\author{Xin Wang}
\email{wangxin579@outlook.com}
\affiliation{Department of Physics, Graduate School of Science, The University of Tokyo, 5-1-5 Kashiwanoha, Kashiwa, Chiba 277-8574, Japan}

\begin{abstract}
    The thermodynamic properties of time-delayed dynamics remain largely unexplored, especially for systems that exhibit asymptotically non-stationary behavior. Here, we investigate heat dissipation in two classes of marginally stable linear time-delayed Langevin dynamics: (i) diffusive criticality, which asymptotically manifests as scaled Brownian diffusion, and (ii) oscillatory criticality, which shows oscillation with diffusive amplitude. By analytical derivations, we find fundamentally different thermodynamic signatures: the average heat dissipation rate asymptotically approaches a constant for diffusive criticality but diverges linearly with oscillations for oscillatory criticality, despite both showing linearly growing variance over time. \red{We discuss in detail how the heat dissipation rate behaves differently as the dynamics asymptotically approaches these two criticality classes from the stable regime.} We \red{also} numerically study the probability distributions of heat dissipation rates for both types of critical dynamics. Our results demonstrate that non-stationary time-delayed dynamics with similar scaling \red{of variance} can yield qualitatively distinct heat dissipation behaviors, depending on the underlying dynamical details. This work provides a concrete foundation for future investigations into thermodynamic properties of general nonlinear time-delayed systems.
\end{abstract}

\maketitle

\section{Introduction}
Time-delayed systems are ubiquitous in natural and engineered systems~\cite{ecosystems,neural1,biology1,biology3,viscous,Feedback,chemical}, where finite signal propagation or processing time introduces nontrivial memory effects. Time-delayed effects may induce complex phenomena, including spontaneous oscillations~\cite{oscillation1,chemical,biology4}, bifurcations~\cite{bifurcation2,bifurcation1}, resonance~\cite{resonance1,resonance2,resonance3}, chaos~\cite{chaos}, synchronization~\cite{synchronization1,synchronization2,synchronization3}, and pattern formation~\cite{pattern1,pattern2}. Investigating these delay-induced effects is crucial for the accurate analysis and effective design of real-world systems.
\red{In the presence of environmental noise, the dynamics of time-delayed systems requires a probabilistic description, such as time-delayed Langevin equations and time-delayed Fokker-Planck equations~\cite{van1992stochastic,coffey2012Langevin,kuchler1992,frank2003,giuggioli2016fokker}. For noisy systems, time-delayed effects may induce nontrivial thermodynamic behaviors due to its inherent non-Markovian nature.} 
 Several studies have revealed non-zero and even negative average heat dissipation in time-delayed Langevin systems in the steady state, highlighting their non-equilibrium character~\cite{munakata1,sarah1,experiment1,experiment2,sarah3,wang1}. Thermodynamic quantities are also discussed using generalized fluctuation theorems and thermodynamic uncertainty relations adapted for time-delayed dynamics~\cite{munakata2,rosinberg1,rosinberg2,rosinberg3, Holubec,sarah2,hasegawa1,TUR,van1}. 
However, current research on the thermodynamic properties of time-delayed systems has primarily focused on stable dynamics (i.e., dynamics that permits a steady state). For systems in non-stationary states, including those exhibiting diffusive or oscillatory behavior under Pyragas time-delayed feedback control~\cite{Pyragas1,Pyragas2,Pyragas3}, the thermodynamic properties have not been comprehensively explored. Even for linear time-delayed Langevin dynamics, a systematic study is lacking on the heat dissipation at marginal stability when the parameters of the system lie on the boundary of the stability conditions~\cite{kuchler1992}.

To address this gap, we study the heat dissipation in two classes of marginally stable linear time-delayed Langevin dynamics: (i) diffusive criticality and (ii) oscillatory criticality. 
We evaluate the average heat dissipation rate through the variance of the system’s position, which is asymptotically determined by the roots of the linear stability equation with zero real parts.
Surprisingly, we find that although both types of criticality exhibit linearly divergent variance in time, they show fundamentally different thermodynamic behaviors: the average heat dissipation rate approaches a constant in the long-time limit for diffusive criticality but diverges linearly in time for oscillatory criticality. 
\red{
Asymptotic analysis also reveals distinct behaviors of the heat dissipation rate as the system parameters approach the criticality boundaries from the stable regime. For systems near diffusive criticality, the steady-state heat dissipation rate asymptotically approaches a constant limiting value consistent with the result for diffusive criticality within a finite timescale, despite the linearly divergent variance of the system's position. For systems near oscillatory criticality, however, the steady-state heat dissipation rate itself diverges asymptotically, requiring an infinitely long time to be attained. Furthermore, the spectral decomposition and distribution of heat dissipation approach a well-defined limiting form near diffusive criticality, but fail to do so near oscillatory criticality.} Our results for these linear systems demonstrate that non-stationary time-delayed dynamics with similar scaling \red{of variance} can produce qualitatively different heat dissipation, depending on the details of the dynamics. This study provides a concrete example describing the thermodynamic properties in time-delayed Langevin systems beyond the stable regime, laying a foundation for broader investigations into nonlinear systems.

This work is structured as follows: Sec.~\ref{section:2} introduces the linear time-delayed Langevin dynamics and specifies the two classes of marginally stable dynamics. Sec.~\ref{section:3} \red{and Sec.~\ref{section:4}} present the core of this work:  Sec.~\ref{section:3} discusses the analytical derivation of the asymptotic average heat dissipation rates for both critical dynamics, alongside numerical results that validate our derivation and explore the full distribution of dissipation. 
\red{Sec.~\ref{section:4} compares the asymptotic behaviors of the average heat dissipation rate for stable dynamics near the stability boundaries of the two distinct criticality classes.}
Finally, Sec.~\ref{section:5} summarizes this work and discusses the future directions.
\vspace{0em}

\section{Marginally stable dynamics of linear time-delayed Langevin systems}\label{section:2}

Consider a class of time-delayed systems governed by the overdamped Langevin equation:
\begin{equation}\label{par_eq}
  \left\{
    \begin{alignedat}{2}
      &\gamma \dot{x}(t) = ax(t) + bx(t-\tau)  + \xi(t) ,
      & (t>t_{\text{init}}), \\ 
      &x(t) = \phi(t),  &\hspace{-2em}(t\in [t_{\text{init}}-\tau ,t_{\text{init}}]),  \\ 
    \end{alignedat}
  \right.
\end{equation}
where $\tau > 0$ is a fixed delay time, $a$ denotes the strength of a linear non-delayed force, and $b$ represents the strength of the linear time-delayed force. The friction coefficient is given by $\gamma$, and $\xi(t)$ is thermal noise satisfying $\left\langle \xi(t)\right\rangle = 0$ and $\left\langle \xi(t)\xi(t')\right\rangle = 2\gamma T \delta (t-t')$, where $T$ is the temperature of the thermal environment. The dynamics is initialized at $t = t_{\text{init}}$ and the past trajectory in $ [t_{\text{init}} - \tau, t_{\text{init}}]$ is described by a predefined function $\phi(t)$. Throughout this work, we set $t_{\text{init}} = 0$ without loss of generality, \red{and we assume that $\phi(t)$ is an absolutely continuous deterministic function and its derivative is integrable over the interval $[-\tau, 0]$.} \red{Physical examples of dynamics Eq.~\eqref{par_eq} include nanomechanical resonators~\cite{rosinberg1,poot2012mechanical} and colloidal particles~\cite{bell2023dynamics,chen2024persistent} under time-delayed feedback control.}

\red{According to Ref.~\cite{kuchler1992}}, the solution to the dynamics in Eq.~\eqref{par_eq} is as follows:
\begin{align}\label{for_sol}
  x(t) &= x_0(t)\phi(0)+
  \red{\frac{b}{\gamma}}\int_{-\tau}^{0} x_0(t-s-\tau)\phi(s)ds \nonumber \\
  & \quad + \red{\frac{1}{\gamma}}\int_{0}^t x_0(t-s)\xi(s) ds,
\end{align}
where the \textit{fundamental solution} $x_0(t)$ is defined as the solution of Eq.~\eqref{par_eq} under the following conditions:
\begin{equation}
    T=0, \ \ \phi(t)=0 \ (t<0),  \ \ \phi(0)=1.
\end{equation}
In Appendix~\ref{appendix:exp}, we show that \red{the fundamental solution can be expanded as} 
\begin{equation}\label{fun_sol_exp} 
    x_0(t)=\sum_{k=-\infty}^{\infty} \dfrac{\red{\gamma}}{\red{\gamma}+b\tau e^{-S_k \tau}}e^{S_kt}  \ \  (t>0), 
\end{equation}
where $\{S_k\} \   (k\in \mathcal{Z})$ correspond to all the roots that satisfy the characteristic equation~\cite{yi1} 
\begin{equation}\label{char_eq}
    \red{\gamma}S_k-a-be^{-S_k\tau}=0.
\end{equation}
This characteristic equation corresponds to the linear stability of the dynamics (Eq.~\eqref{par_eq}).
The distribution of $x(t)$ is Gaussian, with the mean value and variance:
\begin{align}
\langle x(t)\rangle&=x_0(t)\phi(0)+
 \red{\frac{b}{\gamma}} \int_{-\tau}^{0} x_0(t-s-\tau)\phi(s)ds,  \label{mean}  \\
  \langle x^2(t)\rangle &-\langle x(t)\rangle^2=\red{\frac{2T}{\gamma}}\int_0^tx_0^2(s)ds.  \label{var}
\end{align}
\red{Note that the history function $\phi(t)$ enters into the expression only for the mean value, but not for the variance.} Parameters $a,b,\tau$ in Eq.~\eqref{char_eq} determine the distribution of characteristic roots $\{S_k\}$, which further control the asymptotic behavior of the solution $x(t)$ . If $\max\{\text{Re}[S_k]\}<0$, the distribution of $x(t)$ becomes stationary as $t\rightarrow\infty$ and the dynamics is \textit{stable}; if $\max\{\text{Re}[S_k]\}=0$, the variance of $x(t)$ diverges linearly in time and the dynamics is \textit{marginally stable}; if $\max\{\text{Re}[S_k]\}>0$, the dynamics exponentially diverges in time.

There are exactly two classes of marginally stable dynamics~\cite{hayes1950,cooke1982,Hale1977}:
\begin{figure}[tb]
    
  \centering
  \begin{subfigure}[b]{0.35\textwidth}
    \includegraphics[width=\textwidth]{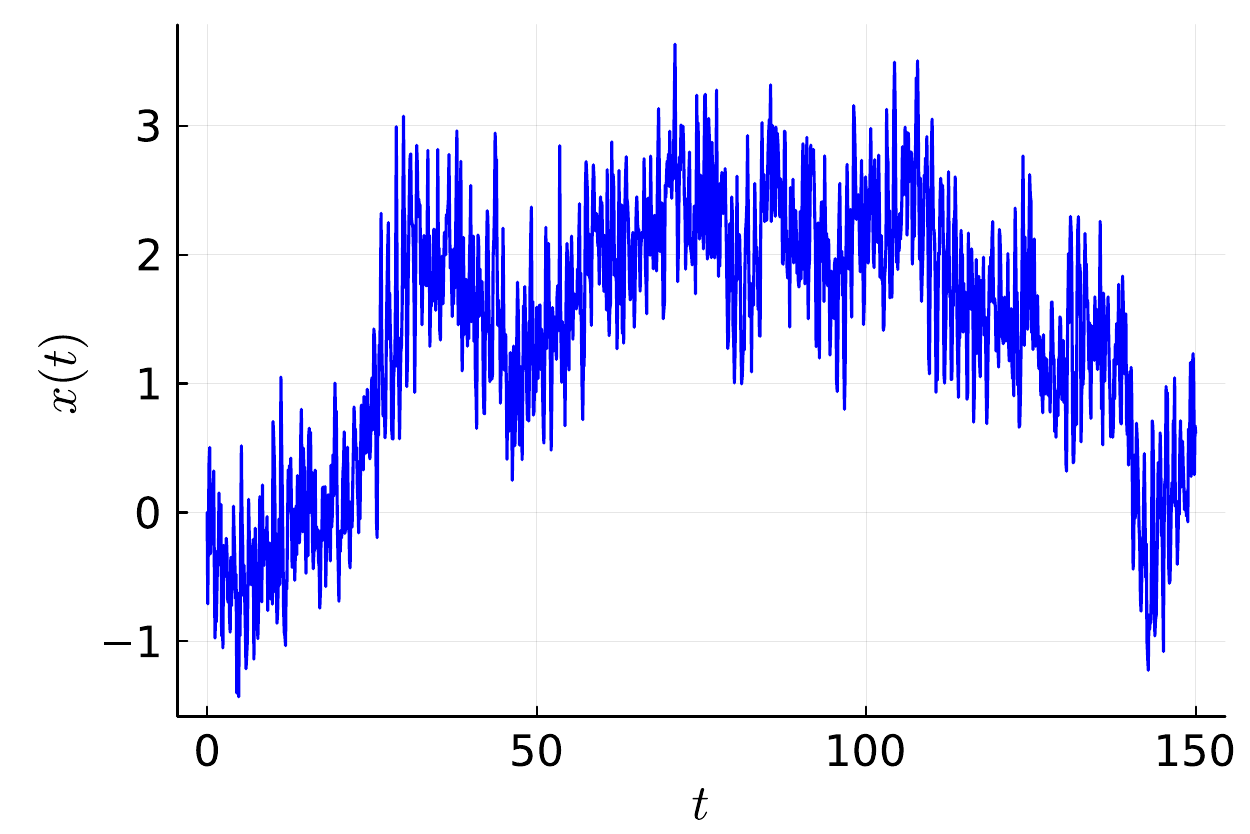}
    \caption{}
    \label{fig:1a}
  \end{subfigure}
  \hfill
  \begin{subfigure}[b]{0.35\textwidth}
    \includegraphics[width=\textwidth]{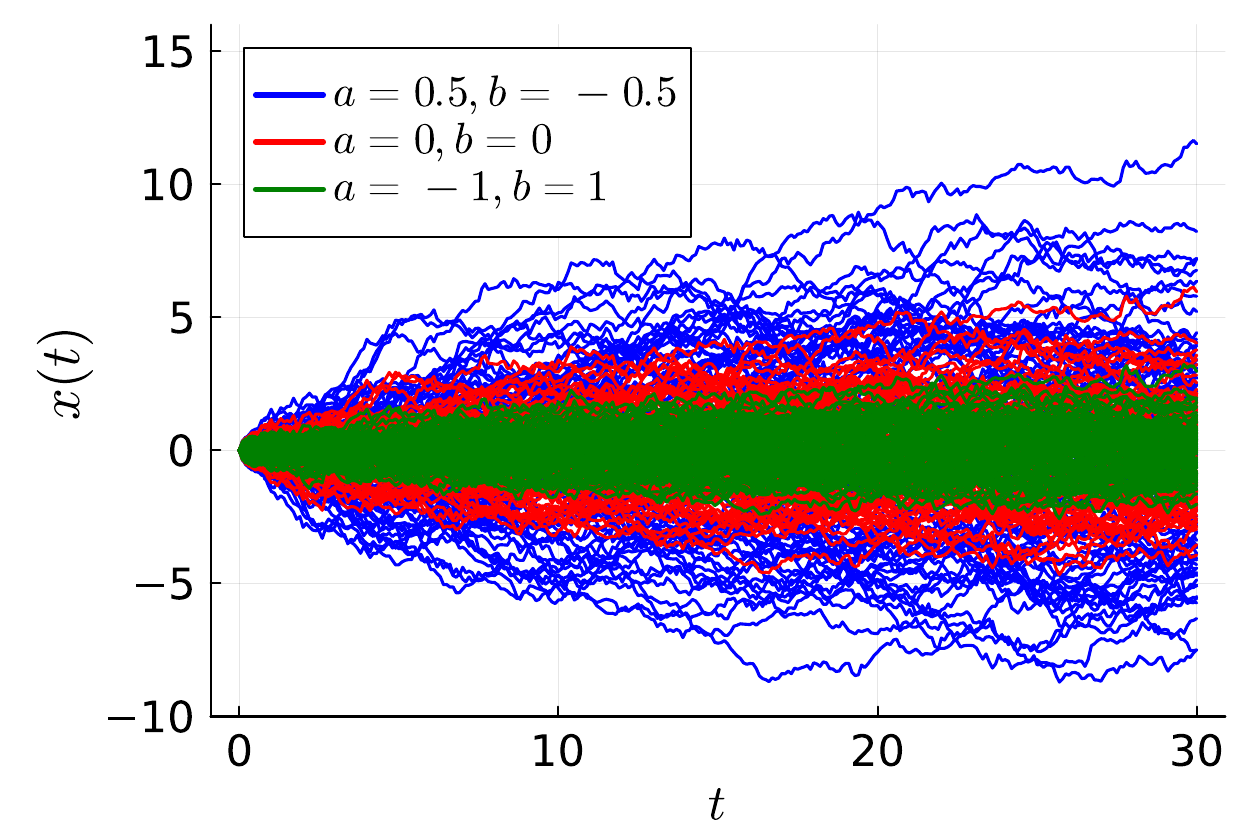}
    \caption{}
    \label{fig:1b}
  \end{subfigure}

  \caption{\raggedright (a) A single trajectory simulation of the dynamics with diffusive criticality. Here $a=-5$, $b=5$, $\tau =1$, $T = 1$, $\red{\gamma=1}$, and $\phi(t)=0  \ \ (t\leq0)$. (b) 100 trajectories of the dynamics with diffusive criticality for each set of parameters. Blue: $a=0.5$ and $b=-0.5$; red: $a=b=0$ (Brownian motion) ; green: $a=-1$ and $b=1$. Here $T=1$, $\red{\gamma=1}$, $\tau=1$, and $\phi(t)=0  \ \ (t\leq0)$. }
  \label{fig:1}
\end{figure}

(i) \textit{Diffusive criticality:} 
When
\begin{equation}\label{cri_con1}
    a+b=0, \ a\tau<\red{\gamma},
\end{equation}
there is exactly one zero root in $\{S_k\}$ and all the others have negative real parts (we exclude $a\tau=\red{\gamma}$ as the coefficients in Eq.~\eqref{fun_sol_exp} diverge). Such critical dynamics corresponds to scenarios of systems under Pyragas control~\cite{Pyragas0, Pyragas1,Pyragas2,Pyragas3}. Under this condition (Eq.~\eqref{cri_con1}), the fundamental solution \red{at large times $t\gg1/\beta_1$} satisfies
\begin{equation}\label{x0_limit01}
    x_0(t)=\dfrac{\red{\gamma}}{\red{\gamma}-a\tau}+O(e^{-\beta_1 t}).
\end{equation}
\red{where  $\beta_1 =\underset{\mathrm{Re}[S_k]<0}{\mathrm{Inf}}
\Big|\mathrm{Re}[S_k]\Big| $ for $\{S_k\}$ in Eq.~\eqref{char_eq} (see Appendix~\ref{section:boundedness} for discussions on asymptotic behavior of the fundamental solution; the same discussion applies to Eq.~\eqref{x0_limit22} at below).} Combining Eq.~\eqref{x0_limit01} with Eqs.~\eqref{mean} and~\eqref{var}, we obtain \red{the behaviors for $t\gg1/\beta_1$}:
\begin{align}
&\langle x(t)\rangle= \frac{\red{\gamma}}{\red{\gamma}-a\tau}\phi(0)-
  \frac{a}{\red{\gamma}-a\tau}\int_{-\tau}^{0} \phi(s)ds+\red{O(e^{-\beta_1 t})},  \ \ \label{mean1}  \\
  &\langle x^2(t)\rangle -\langle x(t)\rangle^2= \frac{2\red{\gamma}T}{(\red{\gamma}-a\tau)^2}\cdot t+C_0+\red{O(e^{-\beta_1 t})} , \ \   \label{var1}
\end{align}
where 
\begin{equation}
    C_0=  \frac{2T}{\red{\gamma}}\int_0^\infty \left[ x_0^2(s)-\left( \dfrac{\red{\gamma}}{\red{\gamma}-a\tau}\right)^2 \right]ds, \label{C_0}
\end{equation}
is a bounded constant \red{(see Appendix~\ref{section:boundedness} for a proof; a similar proof applies to the boundedness of $C_1$ in Eq.~\eqref{C_1} below)}. 
Note that Eq.~\eqref{var1} has been analytically derived in Ref.~\cite{Kobayashi}. The mean displacement $\langle x(t)\rangle$ asymptotically approaches a constant that depends on the historical trajectory $\phi(t)$. The diffusion of the critical dynamics resembles free Brownian motion with a scaling factor $1/(\red{\gamma}-a\tau)^2$ (Fig.~\ref{fig:1a}). A positive time-delayed force ($b=-a>0$) suppresses the diffusion and a negative one ($b=-a<0, \ a\tau<\red{\gamma}$) enlarges the diffusion (Fig.~\ref{fig:1b}).

\begin{figure}[tb]
  \centering
  \begin{subfigure}[b]{0.35\textwidth}
    \includegraphics[width=\textwidth]{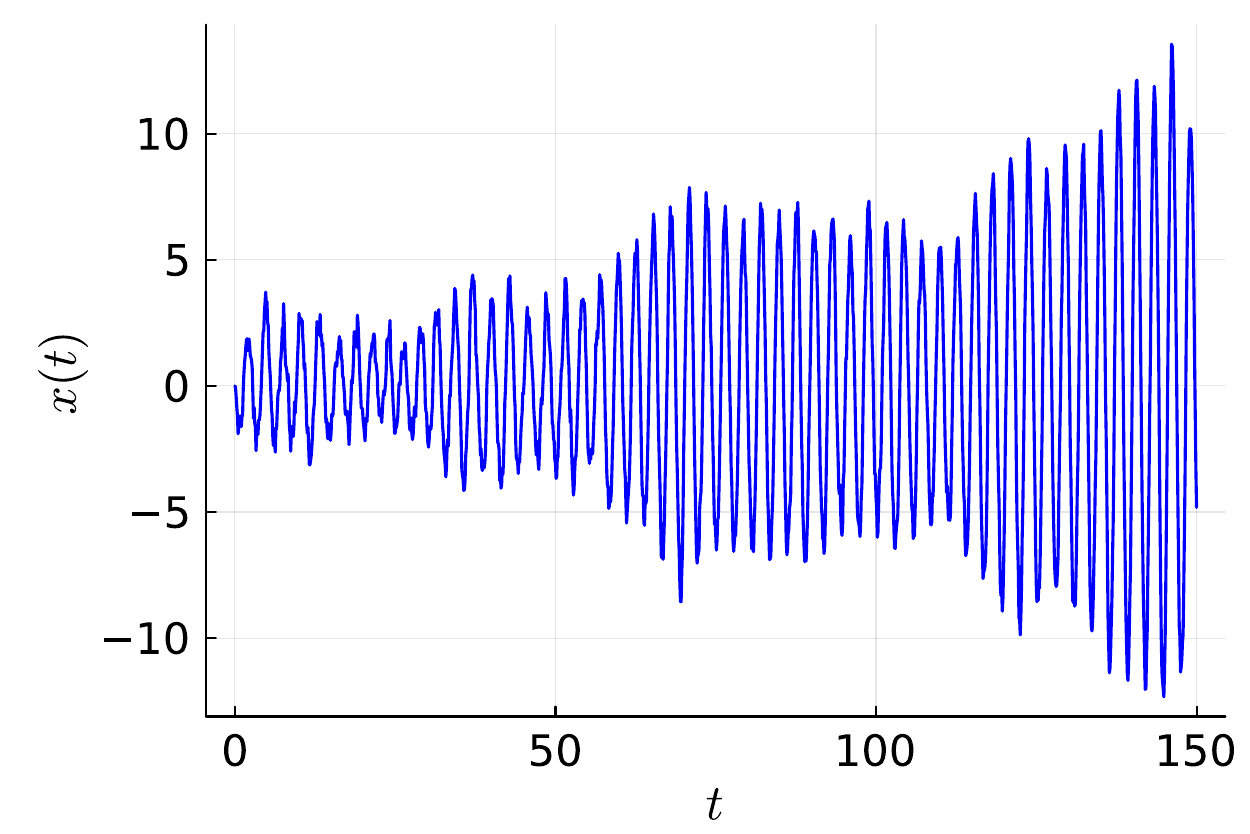}
    \caption{}
    \label{fig:2a}
  \end{subfigure}
  \hfill
  \begin{subfigure}[b]{0.35\textwidth}
    \includegraphics[width=\textwidth]{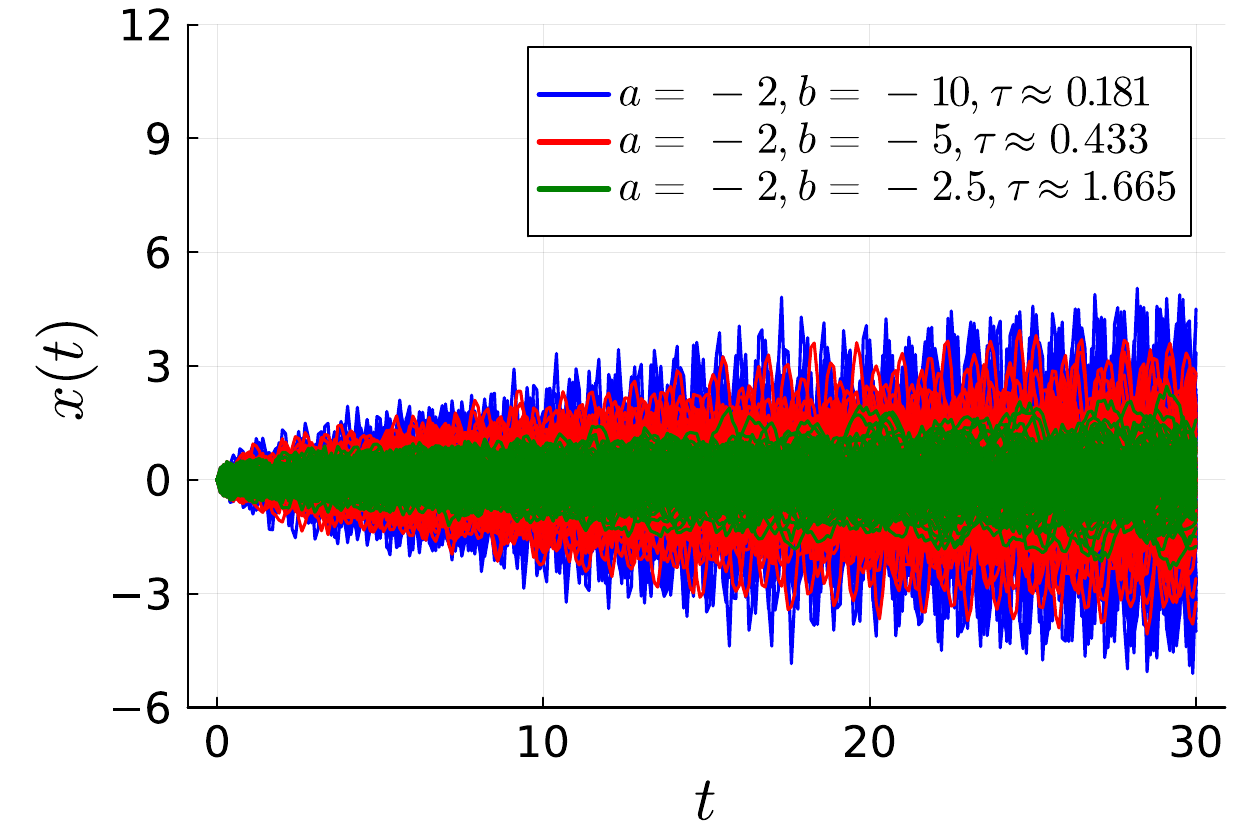}
    \caption{}
    \label{fig:2b}
  \end{subfigure}

  \caption{\raggedright (a) A single trajectory simulation of the dynamics with oscillatory criticality. Here $a=-2$, $b=-3$, $\tau \approx1.029$, $T = 1$, $\red{\gamma=1}$, and $\phi(t)=0  \ \ (t\leq0)$. (b) 100 trajectories of the dynamics with oscillatory criticality for each set of parameters. Blue: $a=-2$, $b=-10$, and $\tau \approx 0.181$; red: $a=-2$, $b = -5$, and $\tau\approx0.433$; green: $a=-2$, $b=-2.5$, and $\tau\approx1.665$. Here $T=1$, $\red{\gamma=1}$, and $\phi(t)=0  \ \ (t\leq0)$.}
  \label{fig:2}
\end{figure}

(ii) \textit{Oscillatory criticality}: 
When
\begin{equation}\label{cri_con2}
    a+b<0, \ a-b>0, \ \text{and}  \ \tau=\red{\tau_c\equiv}\dfrac{\red{\gamma}\arccos\left(-\dfrac{a}{b}\right)}{\sqrt{b^2-a^2}},
\end{equation}
there are exactly two conjugate imaginary roots $\pm i\sqrt{b^2-a^2}/\red{\gamma}$ and all the others have negative real parts. \red{Note that in oscillatory critical dynamics, $\tau$ is uniquely specified by $a$ and $b$.} Such oscillatory critical dynamics shows oscillation with a fluctuating and diffusive amplitude (Fig.~\ref{fig:2a}). The fundamental solution in the long-time limit is
\begin{align}
    x_0(t)&=\frac{\red{\gamma}}{\red{\gamma}+b\red{\tau_c} e^{-i\omega_c\red{\tau_c}}}e^{i\omega_c t} + \frac{\red{\gamma}}{\red{\gamma}+b\red{\tau_c} e^{i\omega_c\red{\tau_c}}}e^{-i\omega_c t}+O(e^{-\beta_1t})  \nonumber \\
   &=\dfrac{2\red{\gamma}\cos(\omega_ct+\theta)}{\sqrt{\red{\gamma^2}+b^2\red{\tau_c}^2-2\red{\gamma}a\red{\tau_c}}}+\red{O(e^{-\beta_1t})}  \label{x0_limit22},
\end{align}
where
\begin{equation}\label{para_x0_limit22}
    \begin{alignedat}{2}
      &\omega_c = \frac{\sqrt{b^2-a^2}}{\red{\gamma}} \ \ \text{and}  
     \ \ \theta=\arctan\left(\frac{\sqrt{b^2-a^2}\cdot \red{\tau_c}}{a\tau-\red{\gamma}} \right).   \\ 
    \end{alignedat}
\end{equation}

The mean displacement \red{at large times $t\gg1/\beta_1$ is} 
\begin{align}
    &\langle x(t)\rangle=\dfrac{2\red{\gamma}}{\sqrt{\red{\gamma^2}+b^2\red{\tau_c}^2-2\red{\gamma}a\red{\tau_c}}} \nonumber \\
    & \cdot\left(\phi(0)\cos(\omega_ct+\theta)+\frac{b}{\red{\gamma}}\int_{-\red{\tau_c}}^0\cos[\omega_c(t-s-\red{\tau_c})+\theta]\phi(s)ds \right)  \nonumber \\ 
    &\ \ \ \ + \red{O(e^{-\beta_1t})},\label{mean2}
\end{align}
and the variance \red{is}
\begin{align}\label{var2}
    &\langle x^2(t)\rangle -\langle x(t)\rangle^2= \nonumber \\
&\dfrac{4\red{\gamma} T \bigg[ t+\frac{1}{2\omega_c}\bigg( \sin[2(\omega_ct+\theta)]-\sin(2\theta) \bigg) \bigg]}{\red{\gamma^2}+b^2\red{\tau_c}^2-2\red{\gamma}a\red{\tau_c}} +C_1+\red{O(e^{-\beta_1t})},
\end{align}
where 
\begin{equation}
    C_1=  \frac{2T}{\red{\gamma}}\int_0^\infty \bigg( x_0^2(s)-\dfrac{4\red{\gamma^2}\cos^2(\omega_cs+\theta)}{\red{\gamma^2}+b^2\red{\tau_c}^2-2\red{\gamma}a\red{\tau_c}}\bigg) ds, \label{C_1}
\end{equation}
is a bounded constant. The mean displacement shows a superposition of oscillations that depends on the historical trajectory $\phi(t)$ and the variance contains a linearly increasing term and an oscillating term. For a fixed $a$, a larger strength of the delayed force $|b|$ leads to oscillations with a shorter period and an amplitude that diffuses faster (Fig.~\ref{fig:2b}).

\section{Heat dissipation of marginal stable dynamics}\label{section:3}
In the following, we analytically derive the asymptotic expressions for the average heat dissipation rate of two types of marginally stable dynamics, along with numerical verification. This section \red{along with Sec.~\ref{section:4}} constitute the main results of our work. 

We define the heat dissipation rate as 
\begin{equation}\label{heat_def}
  dq(t) \equiv [\gamma \dot{x}(t)-\xi(t)]\circ dx(t),
\end{equation}
where $\circ$ denotes the Stratonovich product \cite{sekimoto}. It represents the heat dissipated from the system into the surrounding thermal medium per unit time. \red{A positive sign indicates that the energy is released from the system to the environment.} The heat dissipation rate of steady state for stable solutions has been derived in~\cite{munakata1,sarah1,wang1}. However, those methods are not applicable when calculating the heat dissipation of marginally stable dynamics. In this section, we explicitly calculate the heat dissipation for the two classes of marginally stable dynamics. 

From the definition in Eq.~\eqref{heat_def} and the dynamics described by Eq.~\eqref{par_eq}, the heat dissipation of the system within an infinitesimal time interval $dt$ \red{for $\tau>0$} is given by ($\gamma\equiv1$): 
\begin{align}\label{dq00} 
    &\langle dq(t) \rangle =\left\langle \bigg[\red{\gamma} \frac{dx(t)}{dt}-\xi(t)\bigg]\circ dx(t) \right\rangle\nonumber \\
    &=\left\langle \big[ax(t)+bx(t-\tau)\big]\circ \red{\frac{1}{\gamma}}\big[ ax(t)+bx(t-\tau)+\xi(t) \big]dt  \right\rangle \nonumber \\
    &= \frac{a^2}{\red{\gamma}}\langle x^2(t)\rangle dt+\frac{b^2}{\red{\gamma}}\langle x^2(t-\tau) \rangle dt  \nonumber \\ 
    & \quad \quad+\frac{2ab}{\red{\gamma}}\langle x(t)x(t-\tau)\rangle dt+\frac{aT}{\red{\gamma}}dt, 
\end{align}
where in the last line we use 
\red{
\begin{align}
    &\langle x(t-\tau)\circ\xi(t)dt \rangle= \left\langle \frac{x(t-\tau+dt)+x(t-\tau-dt)}{2} \cdot \xi(t)dt   \right\rangle \nonumber \\
   & = \frac{1}{2\gamma}\int_0^{t-\tau+dt} x_0(t-\tau+dt -s)\langle\xi(s)ds\cdot \xi(t)dt\rangle  \nonumber \\
   &\quad \quad +\frac{1}{2\gamma}\int_0^{t-\tau-dt} x_0(t-\tau-dt -s)\langle\xi(s)ds\cdot \xi(t)dt\rangle \nonumber \\
    &= \frac{1}{2\gamma}\left(\int_0^{t-\tau+dt} x_0(t-\tau+dt-s)\cdot 2\gamma T\delta(t-s) ds\right)\cdot dt \nonumber \\
    &= \frac{1}{2\gamma}\cdot x_0(0)\cdot 2\gamma T\cdot \delta_\tau dt = T\delta_\tau dt.   \quad(\text{since} \  x_0(0)=1)\label{dq}
\end{align}
Here, $dt$ denotes an infinitesimal time interval and $\delta_\tau=1$ when $\tau=0$ and $\delta_\tau=0$ for $\tau>0$. For the time-delayed dynamics, the term $\langle bx(t-\tau)\circ \xi(t)\rangle$ equals zero in the calculation of Eq.~\eqref{dq00} because the system's state at $t-\tau$ is uncorrelated with the noise at $t$~\cite{sarah1}.}   
The correlation $\langle x(t)x(t-\tau)\rangle $ can be calculated through
\begin{align}
    &\frac{1}{2}d\langle x^2(t)\rangle=\langle x(t)\circ dx(t) \rangle \nonumber \\
    &=\frac{a}{\red{\gamma}}\langle x^2(t)\rangle dt+\frac{b}{\red{\gamma}}\langle x(t)x(t-\tau)\rangle dt+\red{\frac{1}{\gamma}}\langle x(t)\circ \xi(t)dt\rangle,
\end{align}
which gives
\begin{equation}
\frac{b}{\red{\gamma}}\langle x(t)x(t-\tau) \rangle=\frac{1}{2}\frac{d\langle x^2(t)\rangle}{dt}-\frac{a}{\red{\gamma}}\langle x^2(t) \rangle
-\frac{T}{\red{\gamma}}.
\end{equation}
Therefore, the average heat dissipation rate is
\begin{equation}\label{dq0}
    \left\langle \frac{dq}{dt} \right\rangle=\frac{b^2}{\red{\gamma}}\langle x^2(t-\tau)\rangle-\frac{a^2}{\red{\gamma}}\langle x^2(t) \rangle + a\frac{d\langle x^2(t)\rangle}{dt} -\frac{aT}{\red{\gamma}}.
\end{equation}
\red{Note that this expression holds for any $\tau>0$ and $t>0$ in spite of the stability of the dynamics.} Below, we discuss the heat dissipation rate of two types of marginally stable dynamics.

\begin{figure}[tb]
  \centering
    \begin{subfigure}[b]{0.35\textwidth}
    \includegraphics[width=\textwidth]{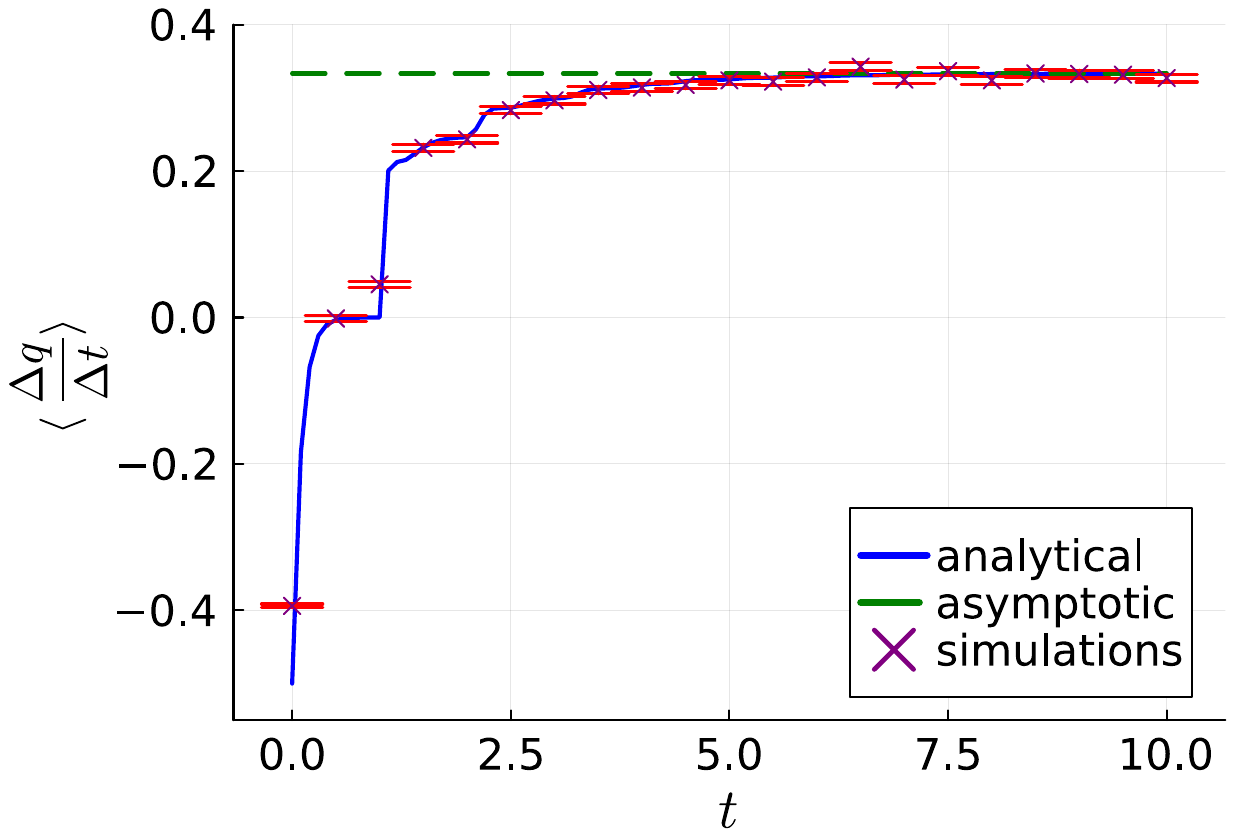}
    \caption{}
    \label{fig:3a}
  \end{subfigure}
  \hfill
  \begin{subfigure}[b]{0.28\textwidth}
    \includegraphics[width=\textwidth]{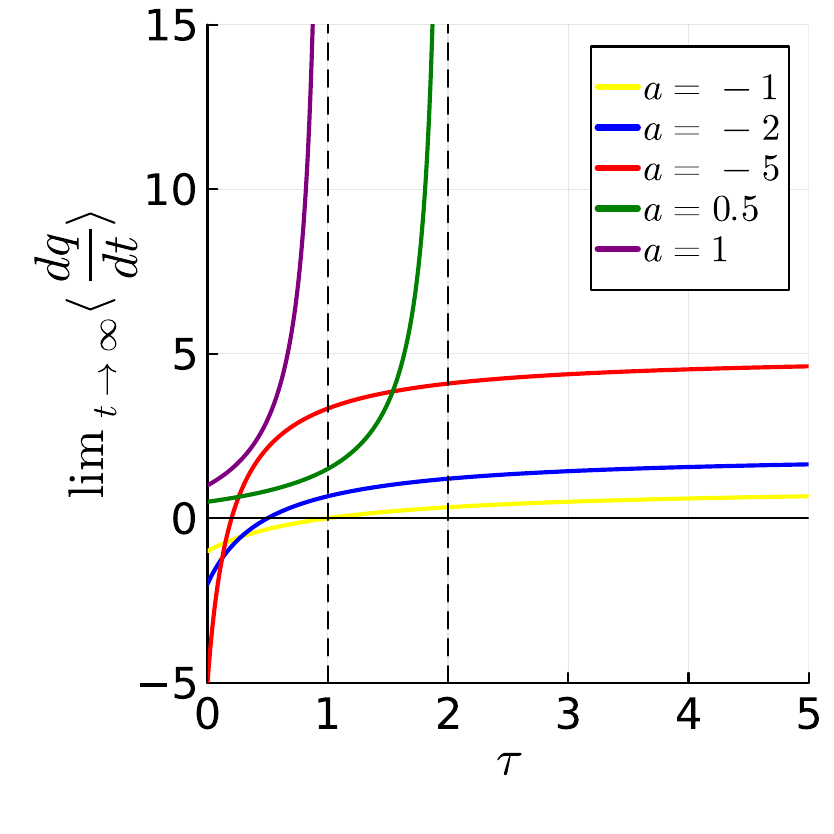}
    \caption{}
    \label{fig:3b}
  \end{subfigure}
  \caption{\raggedright (a) Time evolution of the average heat dissipation rate in diffusive critical dynamics. Solid line: analytical results [calculated by Eqs.~\eqref{var},~\eqref{cri_con1} and ~\eqref{dq0}, where $x_0(t)$ is calculated through the numerical integral of Eq.~\eqref{det_eq}]; dashed line: the asymptotic behavior (Eq.~\eqref{dq1}); crosses: numerical simulations with \red{$50000$} trajectories of precision $dt=10^{-4}$, \red{where $\langle \Delta q/\Delta t\rangle $ is obtained by averaging the values of $\langle dq/dt\rangle$  within a interval $\Delta t=0.1$ centered at each specified time. Red error bars indicate the standard error of the mean.} Here $a = -5$, $b=5$, $\tau = 1$, $T = 0.1$, $\red{\gamma=1}$, and $\phi(t) = 0 \ (t\leq0)$. (b) Average heat dissipation rate of the diffusive critical dynamics in the long-time limit. Here $T=1$ and $\red{\gamma=1}$.}
  \label{fig:3}
\end{figure}

(i) \textit{Diffusive criticality}:

Substituting Eqs.~\eqref{mean1} and~\eqref{var1} into Eq.~\eqref{dq0}, we obtain \red{at large times $t\gg1/\beta_1$:}
\begin{equation}\label{dq1}
    \left\langle \frac{dq}{dt}\right\rangle
    =\frac{aT}{\red{\gamma}}\cdot \dfrac{\red{\gamma}+a\tau}{\red{\gamma}-a\tau}+\red{O(e^{-\beta_1t})}.
\end{equation}
\red{Under the condition $a+b=0$ specified in Eq.~\eqref{cri_con1}, the linearly increasing terms in the second moments of Eq.~\eqref{dq0} cancel out when calculating Eq.~\eqref{dq1}. Consequently, the heat dissipation rate becomes asymptotically constant (Fig.~\ref{fig:3a}) and independent of the history at large times $t \gg 1/\beta_1$.} As shown in Fig.~\ref{fig:3b}, a larger $\tau$ leads to larger heat dissipation rate. The heat dissipation can even become negative for  $a<0$ and $\tau<-\red{\gamma}/a$, despite the diffusive nature of the dynamics. For  $a>0$, the heat dissipation rate diverges as $\tau$ approaches the stability boundary $a\tau=\red{\gamma}$. \red{Note that the average heat dissipation vanishes when $\tau=0$. In this specific limit, the cross-term $\langle bx(t-\tau)\circ \xi(t)\rangle$ in Eq.~\eqref{dq00}, which is zero for $\tau > 0$, now yields a finite contribution of $bT$. This additional term is exactly canceled by other components in Eq.~\eqref{dq00}, resulting in zero dissipation. This unphysical discontinuity in the limit $\tau \rightarrow 0$ (Fig.~\ref{fig:3b}) stems from the idealized assumption of delta-correlated noise in the model.}

\begin{figure}[tb]
  \centering
    \begin{subfigure}[b]{0.35\textwidth}
    \includegraphics[width=\textwidth]{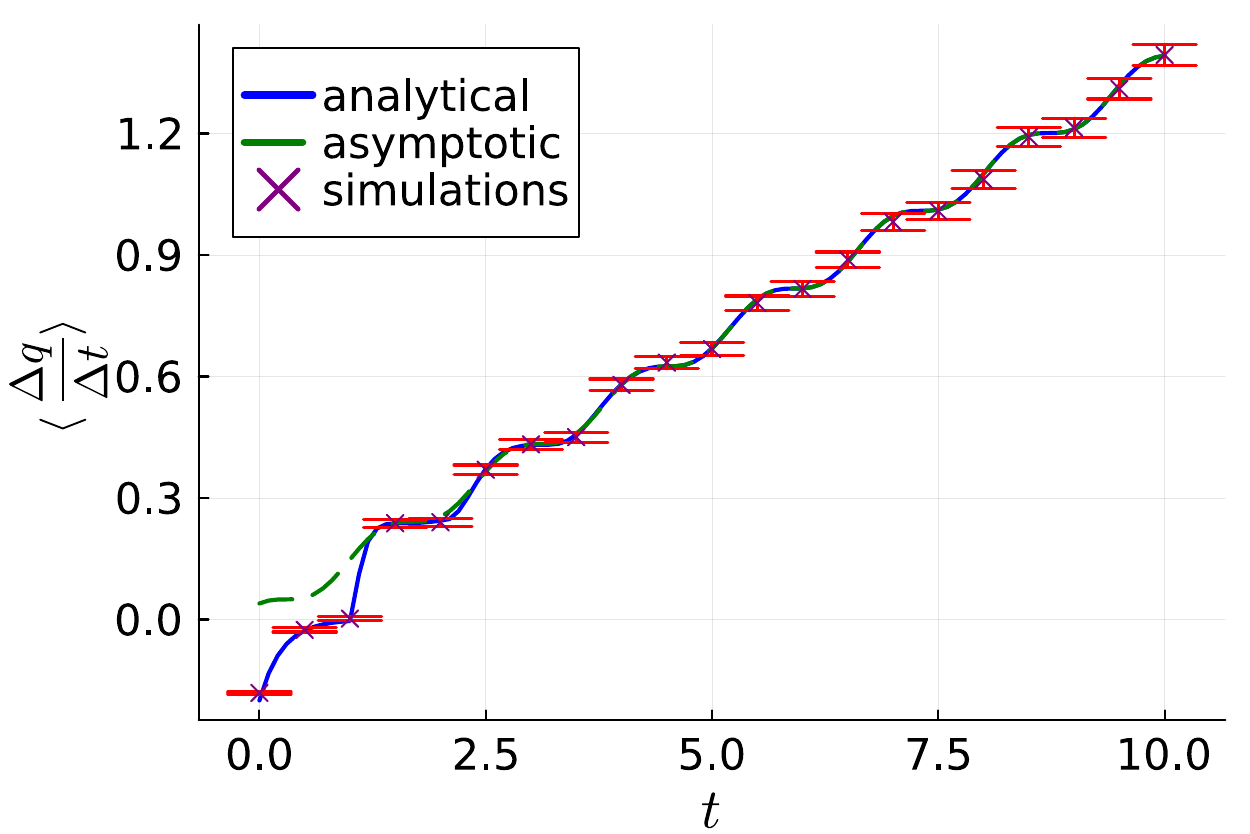}
    \caption{}
    \label{fig:4a}
  \end{subfigure}
  \hfill
  \begin{subfigure}[b]{0.28\textwidth}
    \includegraphics[width=\textwidth]{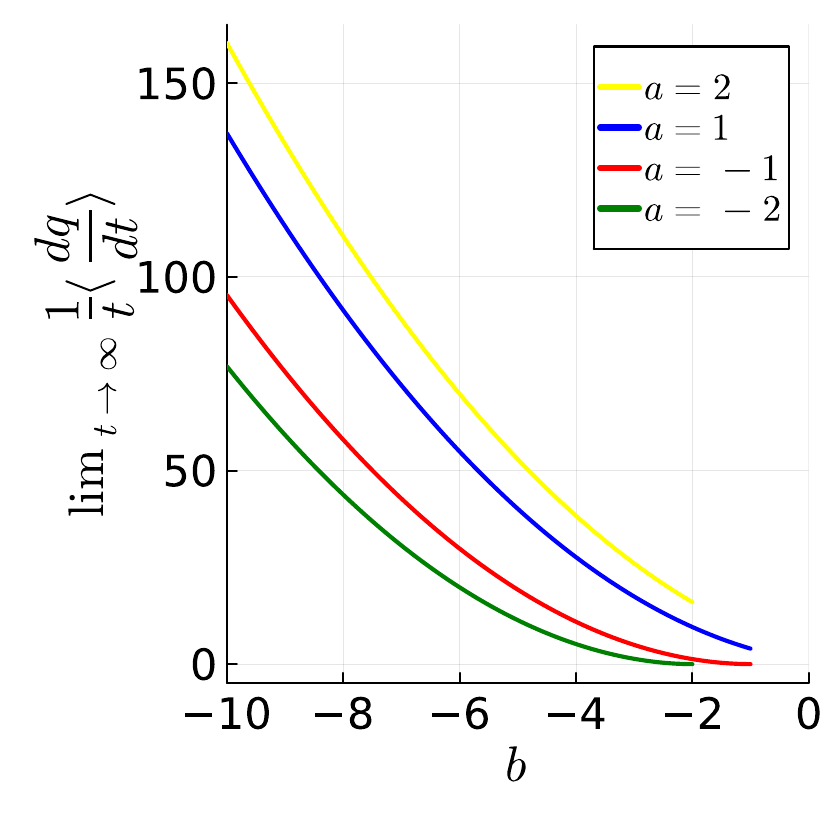}
    \caption{}
    \label{fig:4b}
  \end{subfigure}
  \caption{\raggedright (a) Time evolution of the average heat dissipation rate in oscillatory critical dynamics. Solid line: the analytical result (calculated by Eqs.~\eqref{var},~\eqref{cri_con2}, and~\eqref{dq0}, where $x_0(t)$ is calculated through the numerical integral of Eq.~\eqref{det_eq} ); dashed line: the asymptotic behavior (Eq.~\eqref{dq2}); crosses: numerical simulations with \red{$50000$} trajectories of precision $dt=10^{-4}$, \red{where $\langle\Delta q/\Delta t\rangle$ is obtained by averaging the values of $\langle dq/dt\rangle$  within a interval $\Delta t=0.1$ centered at each specified time. Red error bars indicate the standard error of the mean.} Here $a = -2$, $b=-2.1$, $\tau \approx 4.422$, $T = 0.1$, $\red{\gamma=1}$, and $\phi(t) = 0 \ (t\leq0)$. (b) Divergence speed of the average heat dissipation rate in oscillatory critical dynamics (Eq.~\eqref{lim_q}). Here $b<-|a|$, $T=1$, and $\red{\gamma=1}$.}
  \label{fig:4}  
\end{figure}

(ii) \textit{Oscillatory criticality}:
\red{We set $\phi(t)=0 \ \ (t<0)$ for simplicity, which does not affect the main conclusion.} Substituting Eqs.~\eqref{mean2} and~\eqref{var2} into Eq.~\eqref{dq0}, we obtain \red{at large times $t\gg1/\beta_1$:}

\begin{align}\label{dq2}
    \left\langle \frac{dq}{dt}\right\rangle
    &=\frac{4(b^2-a^2)T}{\red{\gamma^2}+b^2\red{\tau_c}^2-2\red{\gamma}a\red{\tau_c}} \cdot t \  + A_0(t) + B_0  \red{ + O(e^{-\beta_1t})},
\end{align}

where the oscillating term
\begin{align}
    A_0(t) &\equiv \frac{4T}{\red{\gamma^2}+b^2\red{\tau_c}^2-2a\red{\gamma}\red{\tau_c}} \bigg\{ \frac{1}{2\omega_c}\big[  b^2\sin{2\big(\omega_c(t-\red{\tau_c})+\theta\big)}       \nonumber \\ 
    & \ \ -a^2\sin{2(\omega_c t+\theta )} \big]  +a\red{\gamma}\cos{2(\omega_ct+\theta)}\bigg\}
\end{align}
and the constant term
\begin{align}
B_0 &\equiv  \frac{4T}{\red{\gamma^2}+b^2\red{\tau_c}^2-2\red{\gamma}a\red{\tau_c}}\bigg[ -b^2\red{\tau_c}+\frac{(a^2-b^2)\sin{2\theta}}{2\omega_c}+a\red{\gamma} \bigg] \nonumber \\
& \ \ +\frac{(b^2-a^2)C_1}{\red{\gamma}}-\frac{aT}{\red{\gamma}} .   
\end{align}
Note that $A_0(t)+B_0\sim O(1)$; therefore 
\begin{equation}\label{lim_q}
    \lim_{t\rightarrow\infty}\frac{1}{t}\left\langle \frac{dq}{dt}\right\rangle =\frac{4(b^2-a^2)T}{\red{\gamma^2}+b^2\red{\tau_c}^2-2\red{\gamma}a\red{\tau_c}}. 
\end{equation}
\red{A general history function $\phi(t)$ does not influence the asymptotic behavior Eq.~\eqref{lim_q} and only changes the term $A_0(t)$ in Eq.~\eqref{dq2} through the expression of $\langle x(t)\rangle$.} The heat dissipation rate of the dynamics of oscillatory criticality linearly diverges in time accompanied by an oscillation, as shown in Fig.~\ref{fig:4a}. \red{Its divergence rate [Eq.~\eqref{lim_q}] increases with $|b|$ for a fixed $a$ and increases with $a$ for a fixed $|b|$ (Fig.~\ref{fig:4b}).} This shows a clear contrast to the heat dissipation of diffusive critical dynamics, although both dynamics show a variance linearly diverging in time. 

The sharp difference in heat dissipation between the two types of criticality stems from their underlying dynamical details. The dynamics of oscillatory criticality shows oscillation with a linearly growing amplitude on average. \red{For the majority of trajectories, the motion shows an increasing maximum speed, leading to a greater distance covered per oscillation period over time.} The resistance to the noise by such motion ends with linearly divergent heat dissipation in time.
On the other hand, the dynamics of diffusive criticality manifests as a scaled Brownian motion. After a sufficiently long time, the range of diffusion remains local within a fixed interval, which leads to bounded dissipation.

\begin{figure}[tb]
  \centering
  \begin{subfigure}[b]{0.296\textwidth}
    \includegraphics[width=\textwidth]{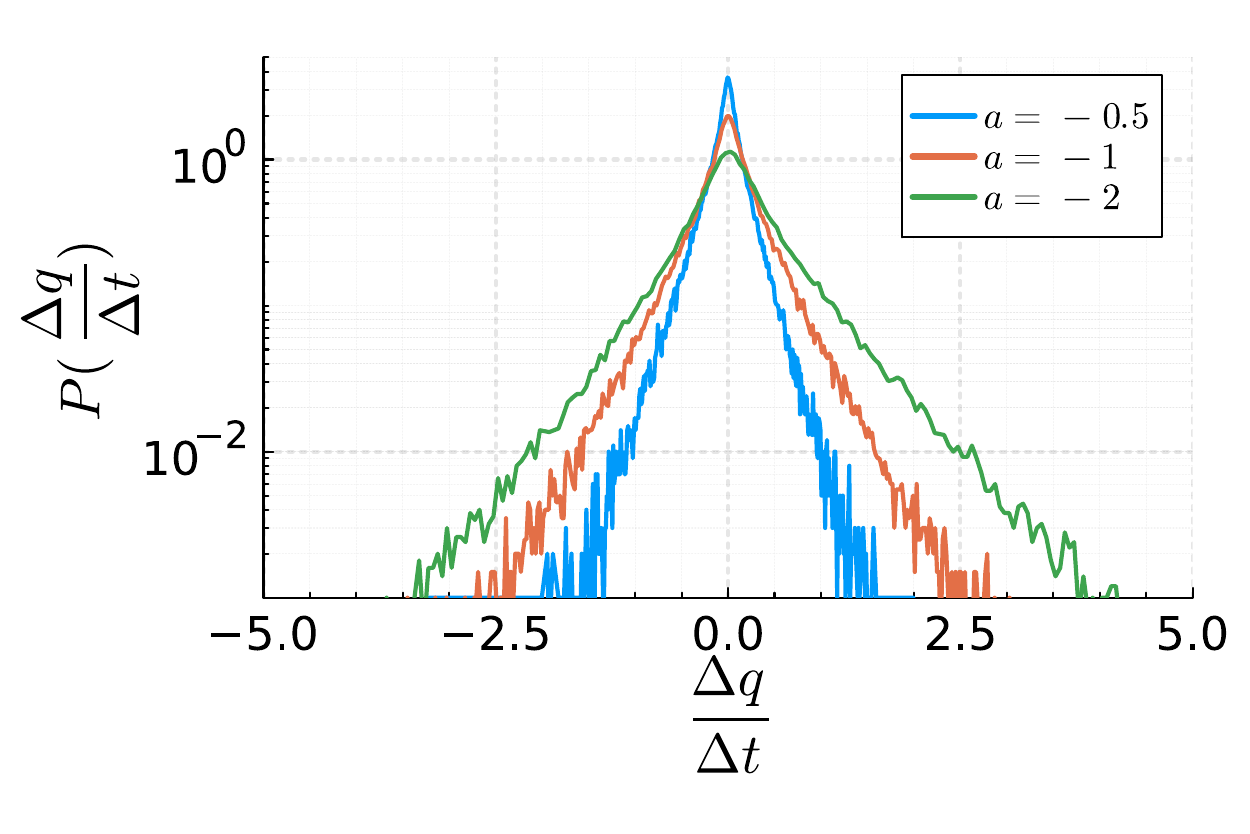}
    \caption{$a=-b<0$}
    \label{fig:5a}
  \end{subfigure}
  \hfill
  \begin{subfigure}[b]{0.296\textwidth}
    \includegraphics[width=\textwidth]{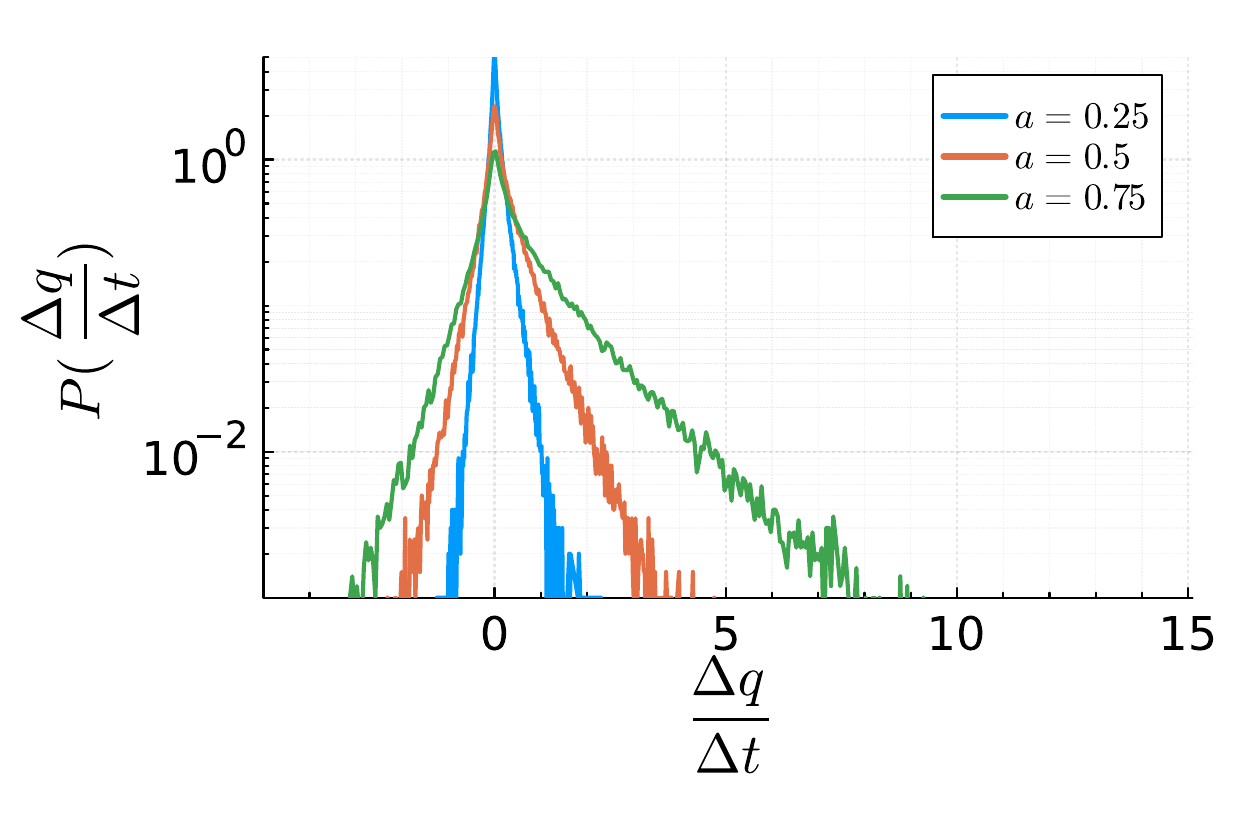}
    \caption{$a=-b>0$, $a\tau<\red{\gamma}$}
    \label{fig:5b}
  \end{subfigure}

  \begin{subfigure}[b]{0.296\textwidth}
    \includegraphics[width=\textwidth]{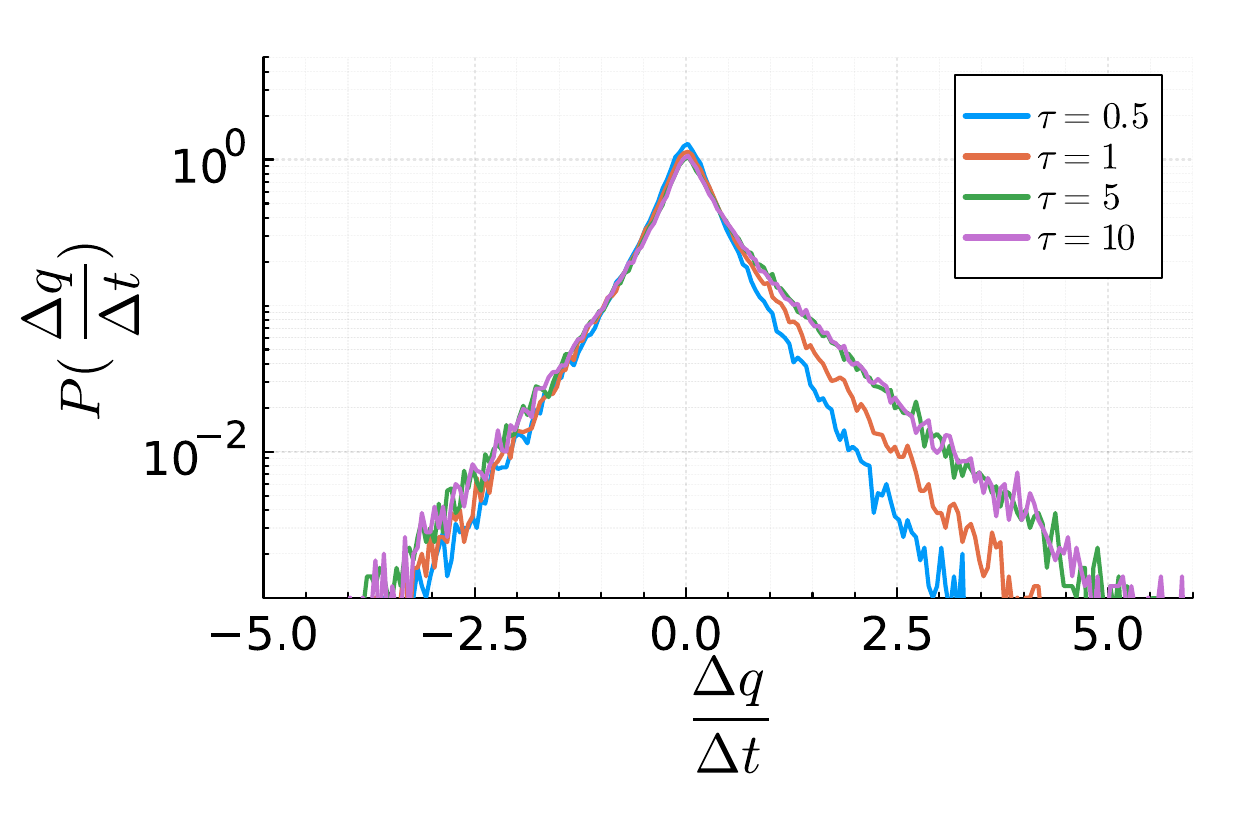}
    \caption{$a=-2$, $b=2$}
    \label{fig:5c}
  \end{subfigure}
    \hfill
  \begin{subfigure}[b]{0.296\textwidth}
    \includegraphics[width=\textwidth]{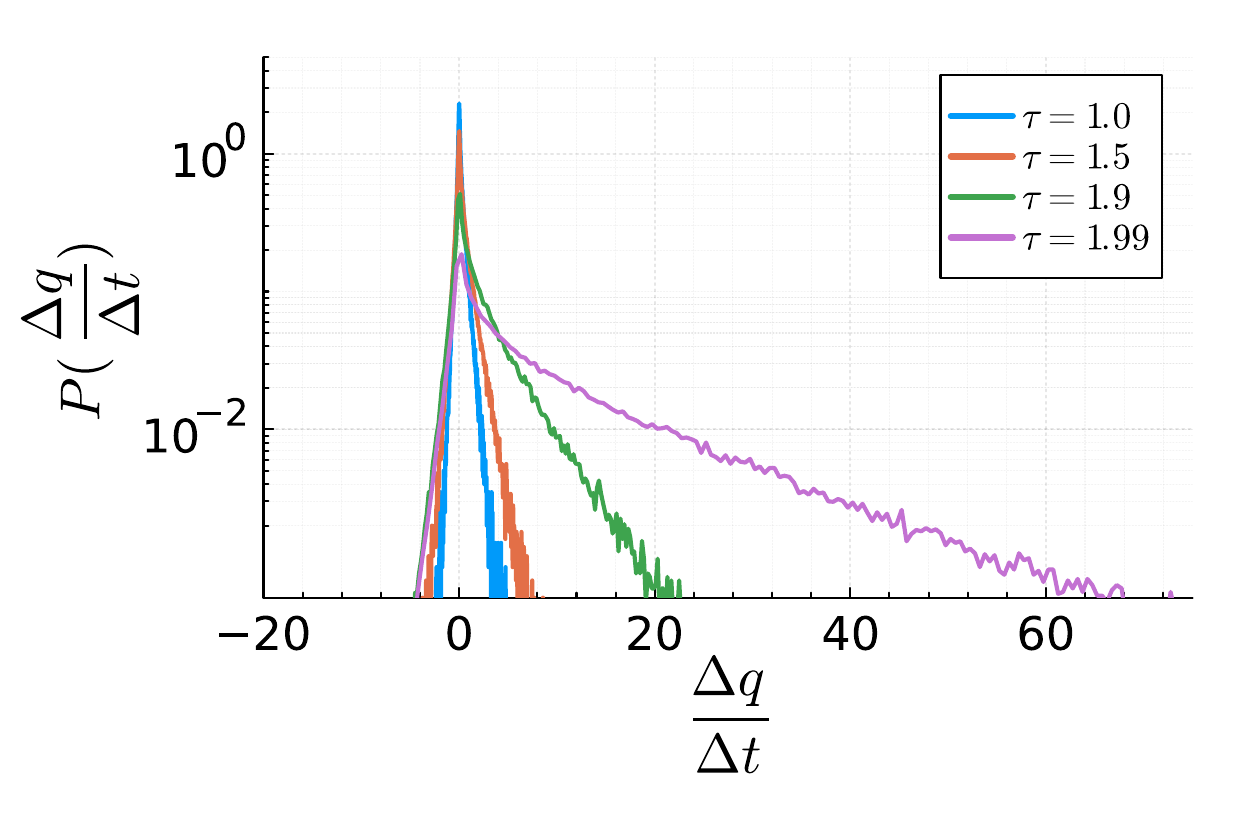}
    \caption{$a=0.5$, $b=-0.5$}
    \label{fig:5d}
  \end{subfigure}
   \caption{\raggedright Probability distribution of the heat dissipation rate $\Delta q/\Delta t$ for diffusive critical dynamics \red{after a sufficiently long time (at $t=70$)}, where each sample of $\Delta q /\Delta t$ is averaged over $\Delta t=\red{1}$ . We take $10^6$ samples with simulation precision $dt=10^{-3}$, and set $T=0.1$ and $\red{\gamma=1}$. (a) Varying $a$ with $a=-b<0$ and $\tau=1$. (b) Varying $a$ with $a=-b>0$ and $\tau=1$. (c) Varying $\tau$ with $a=-2$ and $b=2$. (d) Varying $\tau$ with $a=0.5$ and $b=-0.5$.}
  \label{fig:5}
\end{figure}

\begin{figure}[tb]
  \centering
  \begin{subfigure}[b]{0.325\textwidth}
    \includegraphics[width=\textwidth]{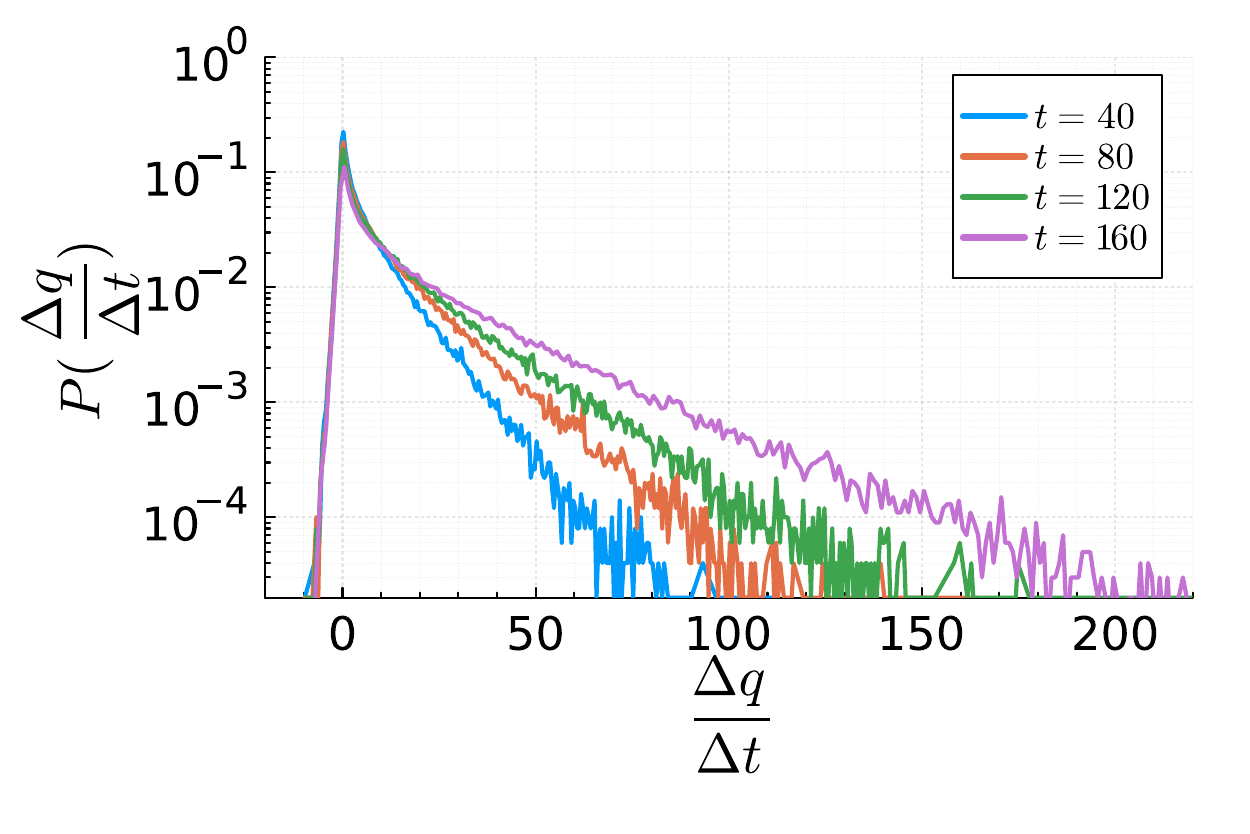}
    \caption{$a=-2$, $b=-3$, $\tau\approx2.301$}
    \label{fig:6a}
  \end{subfigure}

  \begin{subfigure}[b]{0.325\textwidth}
    \includegraphics[width=\textwidth]{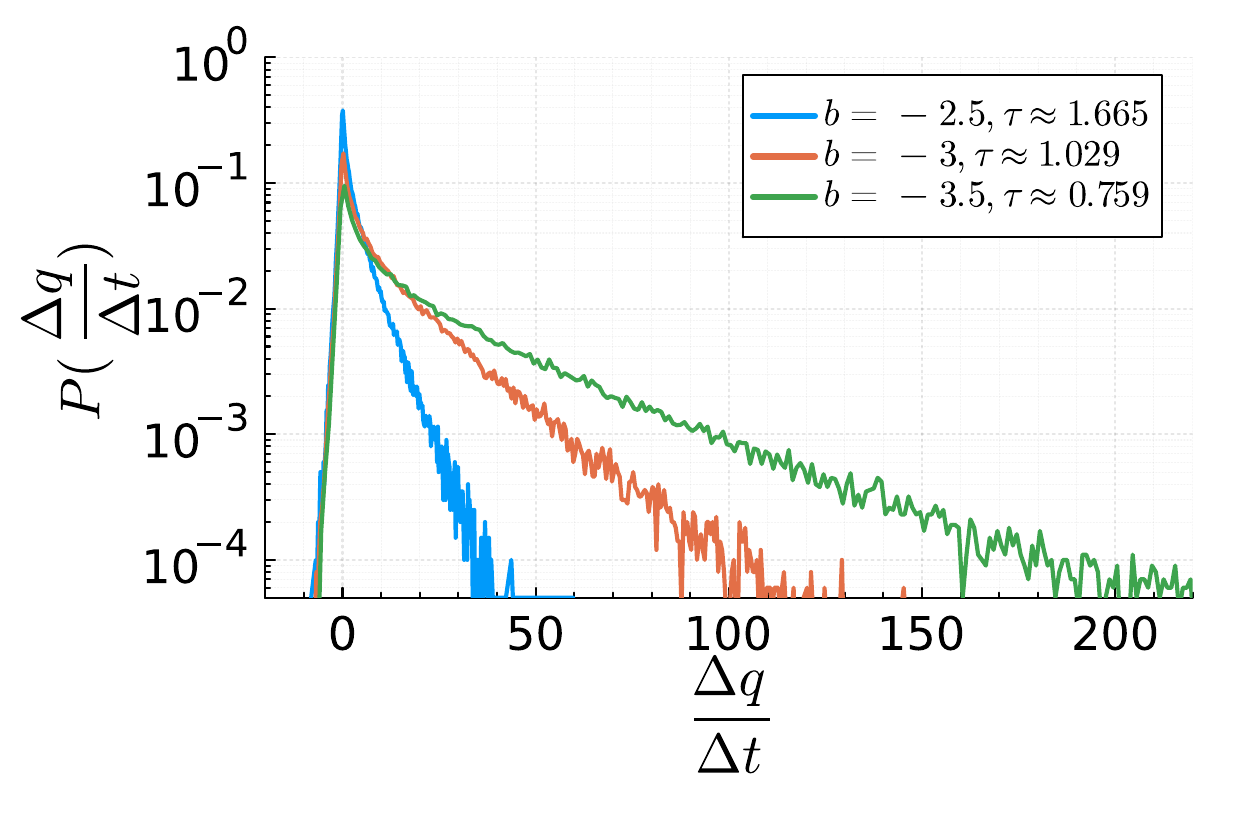}
    \caption{$a=-2$}
    \label{fig:6b}
  \end{subfigure}
    \hfill
  \begin{subfigure}[b]{0.325\textwidth}
    \includegraphics[width=\textwidth]{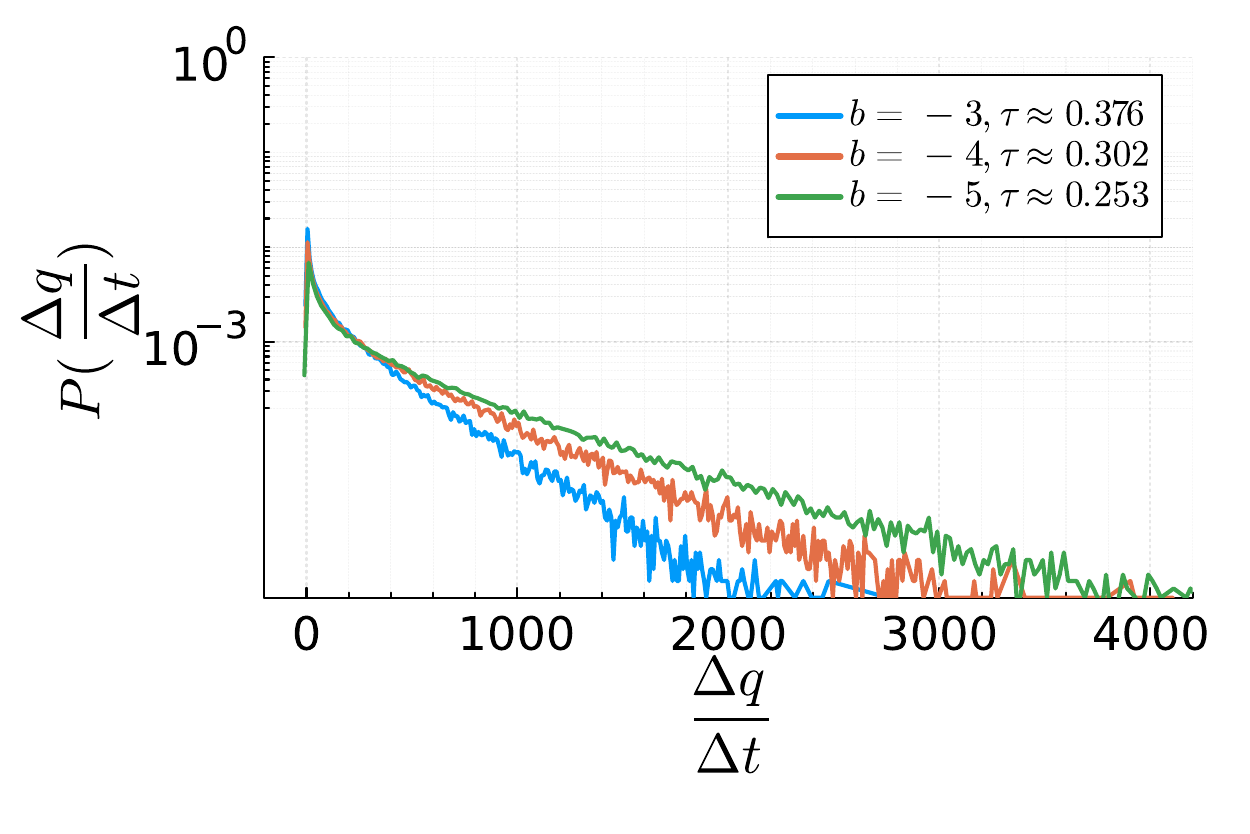}
    \caption{$a=2$}
    \label{fig:6c}
  \end{subfigure}
   \caption{\raggedright Probability distribution of the heat dissipation rate $\Delta q/\Delta t$ for the oscillatory critical dynamics at finite time $t$, where each sample of $\Delta q /\Delta t$ is averaged over $\Delta t=1$. We take $10^{6}$ samples with simulation precision $dt=10^{-3}$, and set $T=0.1$ and $\red{\gamma=1}$. (a) Varying the evolution time $t$ with $a=-2$, $b=-3$, and $\tau\approx2.301$. (b) Varying $b$ with $a=-2$ and $t=70$. (c) Varying $b$ with $a=2$ and $t=70$. }
  \label{fig:6}
\end{figure}

Furthermore, we numerically examine the probability distribution of heat dissipation rate for two types of criticality. The distribution of both criticalities exhibits a single peak and an exponential asymptotic tail. For diffusive criticality, in contrast to the linearly divergent variance of the position, the distribution of heat dissipation converges to an asymptotic one \red{after a sufficiently long time} (Fig.~\ref{fig:5}). Larger  $|a|$ and $|b|$ lead to a broader distribution of heat dissipation (Figs.~\ref{fig:5a} and~\ref{fig:5b}). The distribution also broadens as $\tau$ increases, relaxing to an asymptotic form in the large-$\tau$ limit for $a=-b<0$ (Fig.~\ref{fig:5c}), but spreading without bound as $\tau$ approaches the stability boundary $a\tau=\red{\gamma}$ for  $a=-b>0$ (Fig.~\ref{fig:5d}). For oscillatory criticality, the distribution continues to spread with time without reaching an asymptotic form (Fig.~\ref{fig:6a}). For a fixed $a$, a larger $|b|$ (and thus a smaller $\tau$ according to the marginal-stability conditions) results in a broader distribution (Figs.~\ref{fig:6b} and~\ref{fig:6c}). These features are consistent with earlier analytical results for $\left\langle dq/dt\right\rangle$, but also show how the variance of heat dissipation rate changes with the relevant parameters.

\section{Asymptotic behaviors of heat dissipation rate near the stability boundaries}\label{section:4}
\red{
In this section, we investigate the asymptotic behaviors of the heat dissipation rate as the dynamics [Eq.~\eqref{par_eq}] approaches the two classes of critical dynamics from the stable regime. According to Ref.~\cite{kuchler1992}, the dynamics in Eq.~\eqref{par_eq} is stable if and only if either  
\begin{equation}\label{convergence_condition_over1}
    a+b<0, \ \ a-b\leq0,
\end{equation}
for any $\tau>0$ holds, or
\begin{equation}\label{convergence_condition_over2}
    a+b<0, \ \ a-b>0, \ \ 0<\tau<\frac{\gamma \arccos(-a/b)}{(b^2-a^2)^{1/2}}
\end{equation}
is satisfied. In the stable regime, all characteristic roots in Eq.~\eqref{char_eq} have a negative real part~\cite{kuchler1992}. Therefore, from Eqs.~\eqref{for_sol},~\eqref{mean} and~\eqref{var}, the system asymptotically reaches a steady state in the long-time limit, in which the solution Eq.~\eqref{for_sol} follows a history-independent stationary Gaussian distribution with zero mean and finite variance.  The average heat dissipation rate of the system [Eq.~\eqref{par_eq}] in the steady state can be expressed as~\cite{wang1} (see also~\cite{munakata1,sarah1}):
\begin{equation}\label{heat_stable}
    \langle J\rangle= \frac{aT}{\gamma}+\frac{2T}{\gamma^2}[(a^2+b^2)K(0)+2abK(\tau)],
\end{equation}
where $K(t)\equiv \int_0^\infty x_0(s)x_0(s+t)ds$. The explicit expressions for $K(0)$ and $K(\tau)$ are~\cite{kuchler1992}: 
\begin{align}\label{K_tau}
  K(0)&= \left\{
    \begin{alignedat}{2}
      & \gamma\cdot \dfrac{b\sinh(l\tau/\gamma)-l}{2l(a+b\cosh(l\tau/\gamma))},  \quad \quad  &&(|b|<-a)\\    
      &\dfrac{b\tau-\gamma}{4b}, &&(b=a)  \\ 
      &\gamma\cdot\dfrac{b\sin(l\tau/\gamma)-l}{2l(a+b\cos(l\tau/\gamma))},  &&(b<-|a|)\\ 
    \end{alignedat}
  \right.
  \\
  \label{K_0}
  K(\tau)&= \left\{
    \begin{alignedat}{2}
      & K(0)\cdot \cosh(l\tau/\gamma)-\frac{\gamma\sinh(l\tau/\gamma)}{2l},  \quad  &&(|b|<-a)\\    
      &K(0)-\frac{\tau}{2}, &&(b=a)  \\ 
      &K(0)\cdot \cos(l\tau/\gamma)-\frac{\gamma\sin(l\tau/\gamma)}{2l},  &&(b<-|a|)\\ 
    \end{alignedat}
  \right.
\end{align}
with $l=\sqrt{|a^2-b^2|}$. In the following, we focus on the behaviors of $\langle J \rangle$ in Eq.~\eqref{heat_stable} near the two critical boundaries defined by Eqs.~\eqref{cri_con1} and \eqref{cri_con2}.
}

\red{
First, we fix $a$ and $\tau$, and vary $b$ towards $b=-a$. We take $b=-a-\epsilon_0$, where $|\epsilon_0|\ll|a|$ and according to Eqs.~\eqref{convergence_condition_over1} and~\eqref{convergence_condition_over2}, $\epsilon_0>0$. From the expressions of $K(0)$ and $K(\tau)$ in Eqs.~\eqref{K_0} and~\eqref{K_tau}, we obtain
\begin{align}
    K(0)&= -\frac{\gamma^2}{2\epsilon_0(a\tau-\gamma)}+O(1), \ \ \\
    K(\tau)&=K(0)+\frac{a\tau^2}{2(a\tau-\gamma)}-\frac{\tau}{2}+O(\epsilon_0).
\end{align}
Therefore, using the expression for heat dissipation rate in the stable dynamics:
\begin{align}
    \langle J\rangle&= \frac{aT}{\gamma}+\frac{2T}{\gamma^2}[(a^2+b^2)K(0)+2abK(\tau)] \nonumber \\
    &=\frac{aT}{\gamma}-\frac{4a^2T}{\gamma^2}\left(\frac{a\tau^2}{2(a\tau-\gamma)}-\frac{\tau}{2}  \right)+O(\epsilon_0) \nonumber \\
    &=\frac{aT}{\gamma}\frac{\gamma +a\tau}{\gamma-a\tau}+O(\epsilon_0),
\end{align}
which is consistent with the heat dissipation rate for diffusive criticality [Eq.~\eqref{dq1}]. On the other hand, for oscillatory criticality, we fix $a$ and $b$ and vary $\tau$ towards $\tau_c=\gamma\arccos(-a/b)/(b^2-a^2)^{1/2}$. Let $\tau=\tau_c-\epsilon_1 \ \ (0<\epsilon_1\ll\tau_c)$; then we have
\begin{equation}
    K(0)=\frac{\gamma^2}{(b^2-a^2)\epsilon_1}+O(1), \ \  K(\tau)=-\frac{a}{b}K(0)+O(1),
\end{equation}
and 
\begin{align}\label{asymp_heat_osci}
    \langle J \rangle  &= \frac{T}{\gamma} a + \frac{2T}{\gamma^2} \left[ (a^2 + b^2) K(0) + 2ab \left( -\frac{a}{b} K(0) \right) \right]+O(1)  \nonumber \\
    &=\frac{2T}{\epsilon_1}+O(1),
\end{align}
which shows that the heat dissipation rate in the steady state diverges as the dynamics approaches oscillatory criticality. Note that when the parameters cross the stability boundaries from the stable regime, at least one characteristic root acquires a positive real part. As a result, both the dynamical quantities in Eqs.~\eqref{mean} and~\eqref{var} and the average heat dissipation rate in Eq.~\eqref{dq0} grow exponentially with time.  
}

\red{
To further understand the asymptotic behaviors near criticality, we examine how the dynamical quantities and the average heat dissipation rate respond to infinitesimal shifts of the characteristic root for the two criticality classes. From the characteristic equation~\eqref{char_eq}, we set
\begin{equation}
    F(s) = \gamma s - a - b e^{-s\tau}.
\end{equation}
The marginally stable cases correspond to $F(s_c)=0$, where $s_c$ is the characteristic root with a zero real part. In the case of diffusive criticality, we have $s_c=0$. We consider the dynamics where $b=-a-\epsilon_0 \ \ (\epsilon_0>0)$ and a perturbed characteristic root $s_0=s_c+\Delta s_0$. To first order, we obtain:
\begin{equation}
     \left. \frac{\partial F}{\partial s} \right|_{(s=0,\ b=-a)} \cdot \Delta s_0 - \left. \frac{\partial F}{\partial b} \right|_{(s=0,\ b=-a)} \cdot \epsilon_0 \approx 0,
\end{equation}
which yields
\begin{equation}
     \Delta s_0 \approx -R_0\epsilon_0=O(\epsilon_0)<0,
\end{equation}
where
\begin{equation}
    R_0\equiv \frac{1}{\gamma - a\tau}.
\end{equation}
While for oscillatory criticality,  we have $s_c=\pm i\omega_c$.  Without loss of generality, let $s_c = i\omega_c$. For the dynamics with $\tau = \tau_c - \epsilon_1 \ (\epsilon_1 > 0)$, we consider a perturbed characteristic root $s_1 = s_c + \Delta s_1$ and define $\Delta s_1 = \mathrm{Re}[s_1] + i(\omega_2 - \omega_c)$. We then obtain to first order: 
\begin{equation}
 \left. \frac{\partial F}{\partial s} \right|_{(i\omega_c, \tau_c)}\cdot\Delta s_1- 
\left. \frac{\partial F}{\partial \tau} \right|_{(i\omega_c, \tau_c)}\cdot\epsilon_1\approx0,
\end{equation}
which leads to
\begin{align}
  \Delta s_1 &\approx \frac{\left( \frac{\partial F}{\partial \tau} \right) \epsilon_1}{\frac{\partial F}{\partial s}}= \frac{b i\omega_c e^{-i\omega_c \tau_c}\cdot \epsilon_1}{\gamma + b\tau_c e^{-i\omega_c \tau_c}}\ \nonumber \\
  &= \frac{(-\gamma \omega_c^2 - i a \omega_c) \epsilon_1}{(\gamma - a\tau_c) + i\gamma \omega_c \tau_c}.
\end{align}
In the last line, we have used $b e^{-i\omega_c \tau_c} = i\gamma \omega_c - a$, which follows from $F(s=i\omega_c, \  \tau=\tau_c)=0$. 
This shows
\begin{equation}
    \mathrm{Re}[s_1]\approx -R_1\epsilon_1=O(\epsilon_1)<0,
\end{equation}
where
\begin{equation}
    R_1\equiv \frac{\gamma^2 \omega_c^2}{(\gamma - a\tau_c)^2 + \gamma^2 \omega_c^2 \tau_c^2}.
\end{equation}
Therefore, for stable dynamics close to either type of criticality, it follows from Eqs.~\eqref{mean},~\eqref{var} and~\eqref{x_0_bound2} that  
\begin{equation}\label{x0_dt_asymp}
    \frac{d}{dt}\langle x^2(t)\rangle =\frac{2T}{\gamma}x_0^2(t)=O(e^{-2\varepsilon t})+O(e^{-(\varepsilon+|\mathrm{Re}[\nu]|)t})+\cdots,
\end{equation}
where $\varepsilon=R_0\epsilon_0$ (or $\varepsilon=R_1\epsilon_1$), and $\nu=\nu_0$ (or $\nu=\nu_1$) is the characteristic root with the second largest real part  for the dynamics close to  diffusive (or oscillatory) criticality. This shows that in the limit $\epsilon_0\rightarrow0$ or $\epsilon_1\rightarrow0$, the dynamics close to either diffusive or oscillatory criticality requires asymptotically infinite time (of order $O(1/\epsilon_0)$ or $O(1/\epsilon_1)$, respectively) to achieve the steady state. However, the rate of convergence for the average heat dissipation rate differs between the two classes of criticality. From Eqs.~\eqref{dq0} and~\eqref{x0_dt_asymp}, for the dynamics close to diffusive criticality, we have 
\begin{align}\label{asymp_anal_dq1}
    &\frac{d}{dt}\left\langle\frac{dq(t)}{dt} \right\rangle =\frac{1}{\gamma}\frac{d}{dt}\bigg( b^2[\langle x^2(t-\tau)\rangle-\langle x^2(t)\rangle]  \nonumber \\
    &\quad \quad \quad +(b^2-a^2)\langle x^2(t)\rangle+a\gamma\frac{d\langle x^2(t)\rangle}{dt}-aT\bigg)
    \nonumber \\  &=O(R_0\epsilon_0e^{-2R_0\epsilon_0t}) 
    +O([R_0\epsilon_0+|\nu_0|]e^{-(R_0\epsilon_0+|\mathrm{Re}[\nu_0]|)t})+\cdots.
\end{align}
In the last line of Eq.~\eqref{asymp_anal_dq1}, we have used 
\begin{align}
    &b^2[\langle x^2(t-\tau)\rangle-\langle x^2(t)\rangle] =\frac{2Tb^2}{\gamma}\int_{t-\tau}^tx_0^2(s)ds \nonumber \\
    &\quad =O(b^2\tau\cdot e^{-2R_0\epsilon_0 t})+\cdots, \\
    &b^2-a^2=2a\epsilon_0+O(\epsilon_0^2).
\end{align}
On the other hand, for the dynamics close to the oscillatory criticality, we have 
\begin{equation}\label{asymp_anal_dq2}
    \frac{d}{dt}\left\langle\frac{dq(t)}{dt} \right\rangle  = O(e^{-2R_1\epsilon_1t})+O(e^{-(R_1\epsilon_1+|\mathrm{Re}[\nu_1]|)t})+\cdots,
\end{equation}
because in this case $b^2-a^2\sim O(1)$ and $(b^2-a^2)\cdot d\langle x^2(t)\rangle/dt$ accounts for the leading-order contribution in Eq.~\eqref{asymp_anal_dq2}. Therefore, in the limit $\epsilon_0\rightarrow0$ or $\epsilon_1\rightarrow0$, the heat dissipation rate of stable dynamics near diffusive criticality approaches a constant on a finite time scale of order $O(1/|\mathrm{Re}[\nu_0]|)$, whereas that near oscillatory criticality requires an asymptotically infinite time of order $O(1/\epsilon_1)$ to reach the specific steady-state value given by the expression Eq.~\eqref{heat_stable}  for stable dynamics.
}

\red{
We also investigate the spectral decomposition of the steady-state heat dissipation rate as the parameters asymptotically approaches the two critical boundaries. According to the Harada-Sasa equality, the steady-state heat dissipation rate can be decomposed as follows~\cite{harada-sasa1}:
\begin{equation}\label{hs}
\left\langle J \right\rangle = \gamma \left\{ v_s^2 + \int_{-\infty}^{\infty}\left[\tilde{C}_v(\omega)-
2T\tilde{R}'_v(\omega)\right] \frac{d\omega}{2\pi}\right\}.   
\end{equation}
Here, $v_s$ is the steady-state velocity,  $\tilde{C}_v(\omega)$ is the velocity power spectral density and $\tilde{R}_v'(\omega)$ is the real part of linear response function. The integrand $\tilde{C}_v(\omega)-
2T\tilde{R}'_v(\omega)$ quantifies the violation of the fluctuation-response relation, which vanishes for all $\omega$ in equilibrium systems but is generally non-zero in non-equilibrium steady states~\cite{kubo,harada-sasa2}. The Harada-Sasa equality has been shown to remain valid for time-delayed Langevin systems~\cite{ito,wang1}. 
}
\begin{figure}[tb]
    
  \centering
  \begin{subfigure}[b]{0.35\textwidth}
    \includegraphics[width=\textwidth]{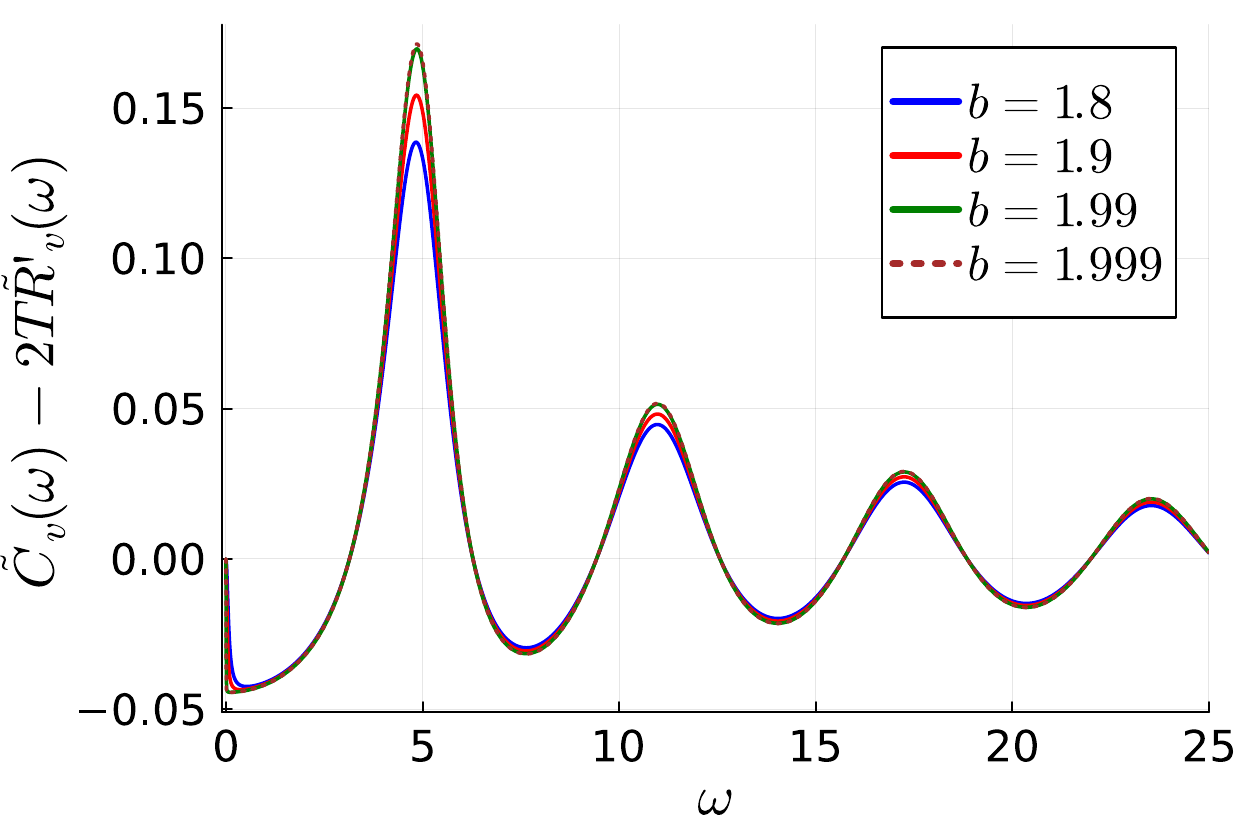}
    \caption{$a=-2$}
    \label{fig:7a}
  \end{subfigure}
  \hfill
  \begin{subfigure}[b]{0.35\textwidth}
    \includegraphics[width=\textwidth]{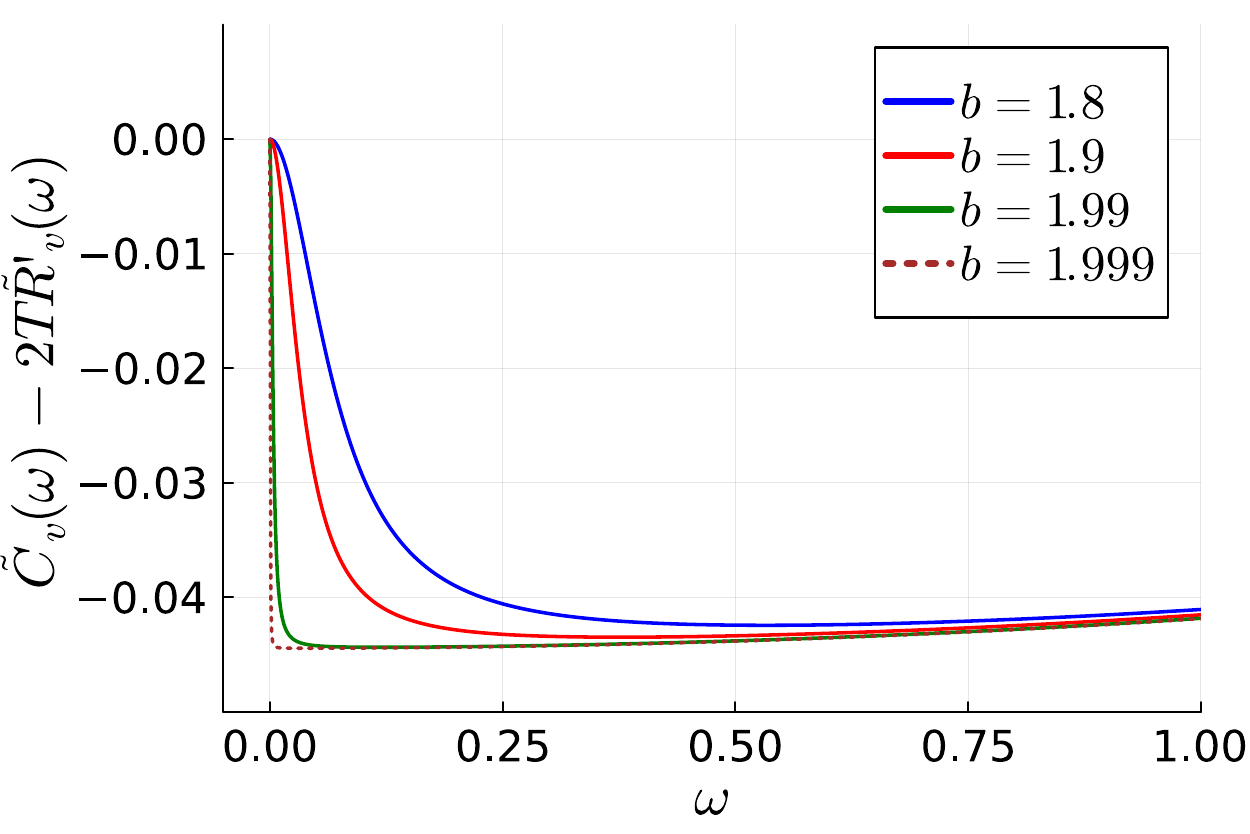}
    \caption{$a=-2$}
    \label{fig:7b}
  \end{subfigure}
 \hfill
  \begin{subfigure}[b]{0.35\textwidth}
    \includegraphics[width=\textwidth]{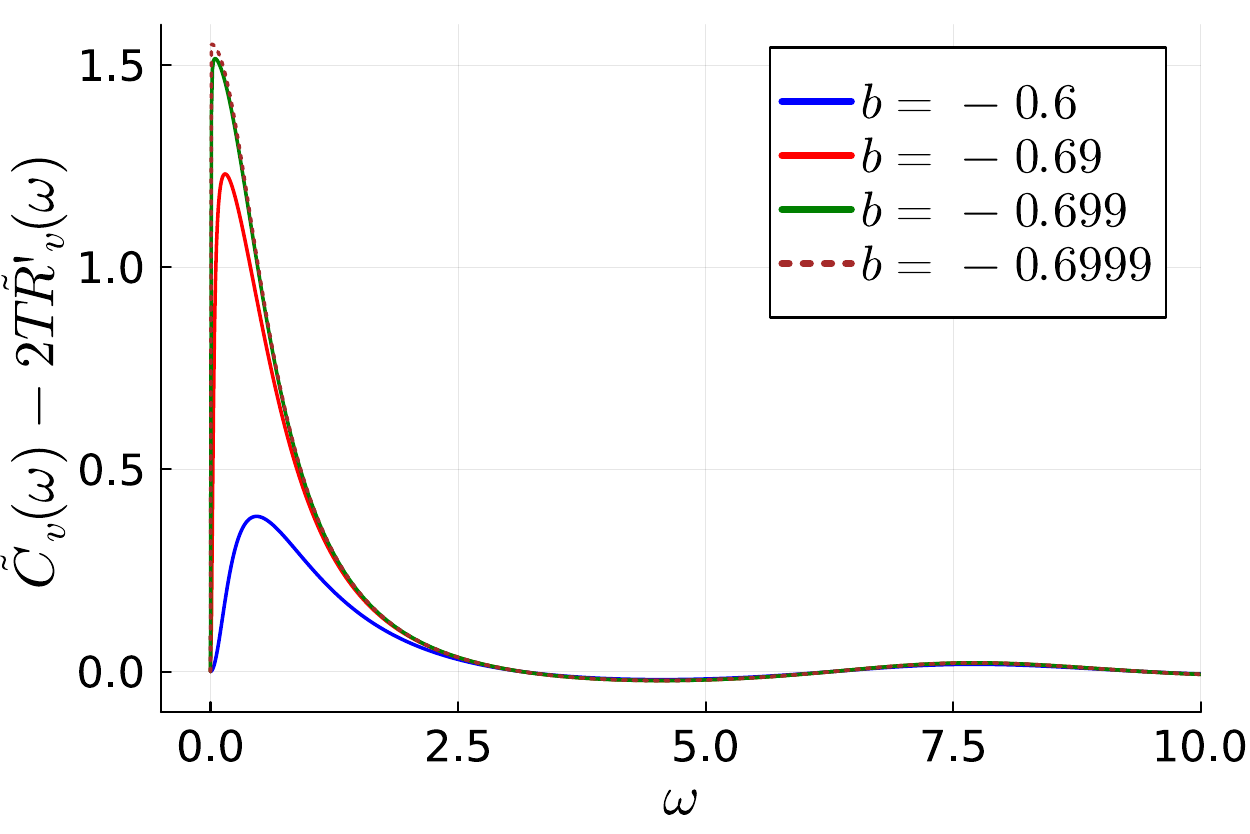}
    \caption{$a=0.7$}
    \label{fig:7c}
  \end{subfigure}

  \caption{\raggedright \red{Spectral decomposition of the heat dissipation rate in the steady state in linear time-delayed Langevin equation [Eq.~\eqref{par_eq}] near the diffusive criticality. Note that (a) and (b) show the same curves with different range of $\omega$.
  Here, we fix $\tau=1$, $T=0.1$, and $\gamma=1$.}   }
  \label{fig:7}
\end{figure}

\begin{figure}[tb]
  \centering
  \begin{subfigure}[b]{0.35\textwidth}
    \includegraphics[width=\textwidth]{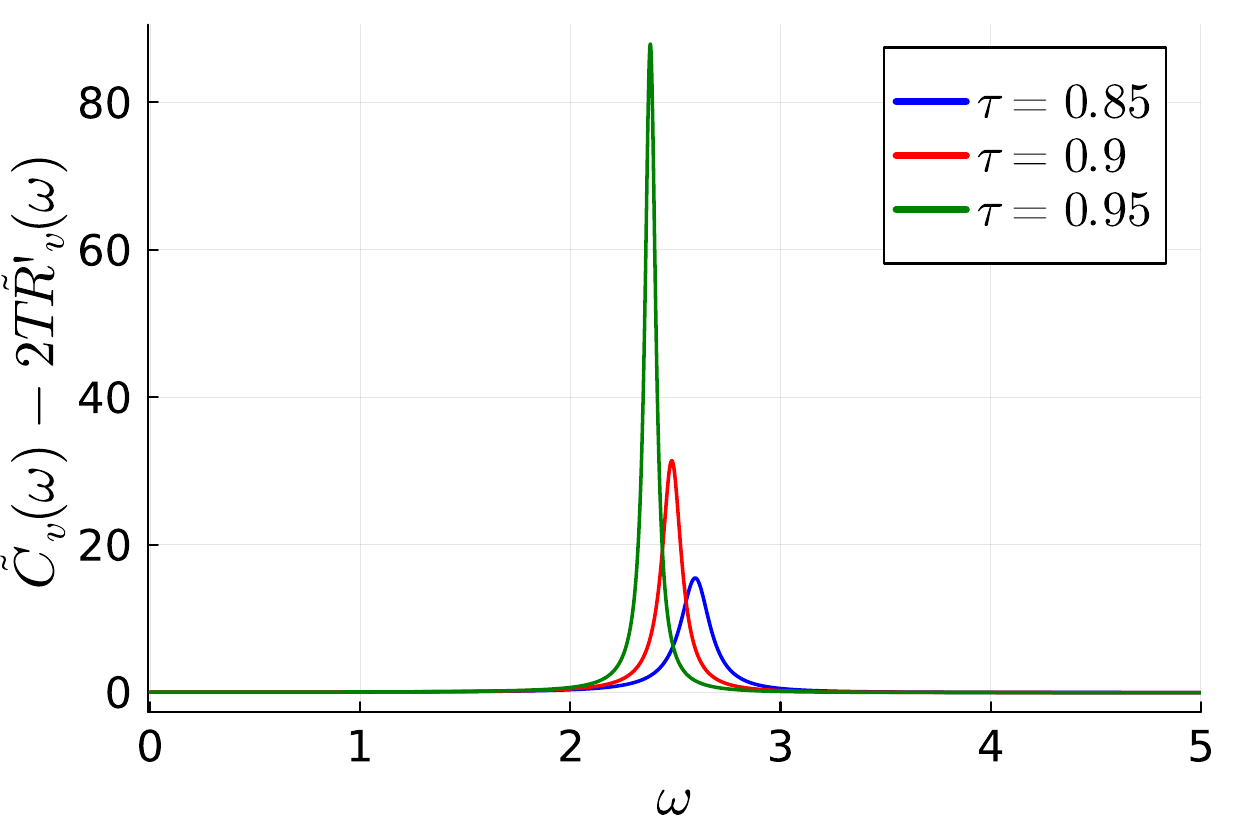}
    \caption{$a=-2$}
    \label{fig:8a}
  \end{subfigure}
  \hfill
  \begin{subfigure}[b]{0.35\textwidth}
    \includegraphics[width=\textwidth]{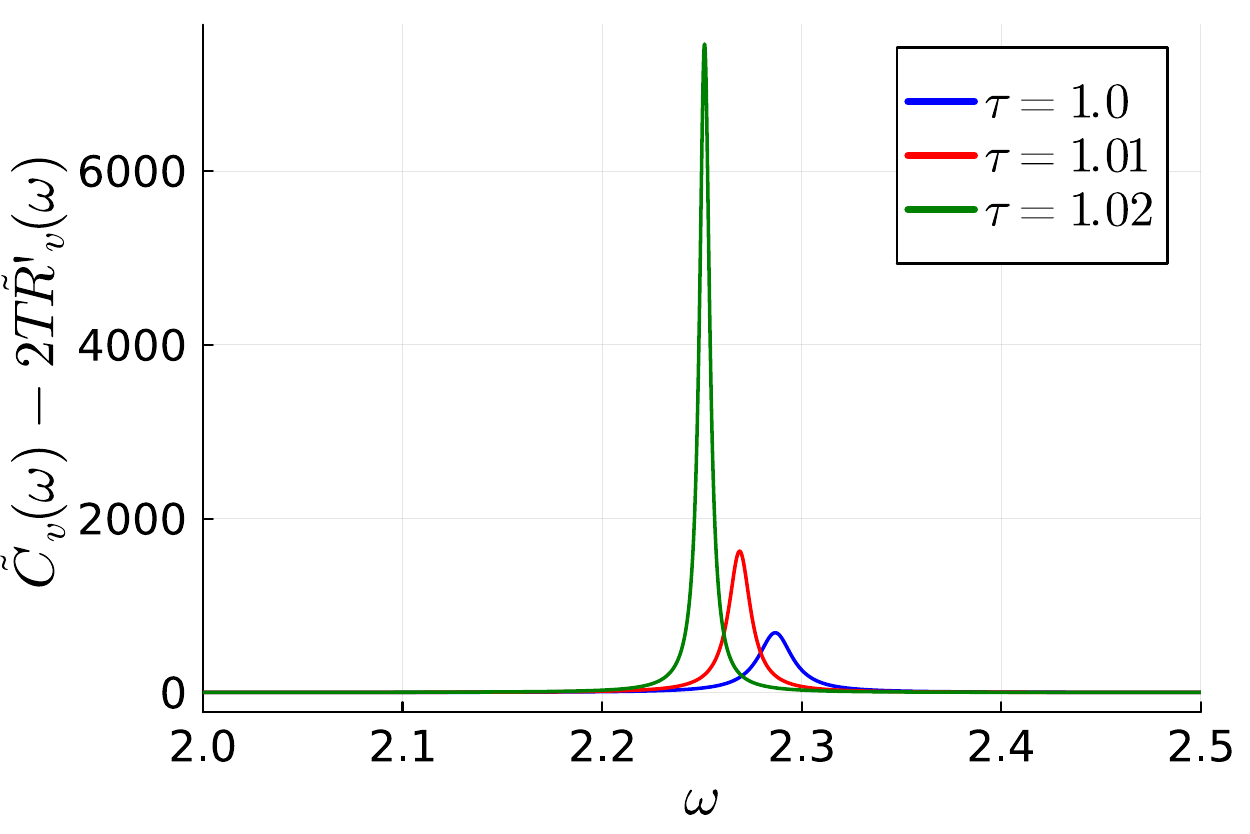}
    \caption{$a=-2$}
    \label{fig:8b}
  \end{subfigure}

  \caption{\raggedright \red{Spectral decomposition of the heat dissipation rate in the steady state in linear time-delayed Langevin equation [Eq.~\eqref{par_eq}] near the oscillatory criticality.
  Here, we fix $a=-2$, $b=-3$, $T=0.1$ and $\gamma=1$. Note that in these cases, $\tau_c\approx1.0288$ and $\omega_c\approx2.236$.}}
  \label{fig:8}
\end{figure}

\red{
In particular, for the linear time-delayed dynamics [Eq.~\eqref{par_eq}] under the stability condition Eq.~\eqref{convergence_condition_over1} or~\eqref{convergence_condition_over2}, the steady-state velocity is $v_s=0$. The integrand in the spectral decomposition of the steady-state heat dissipation rate is given by~\cite{wang1}:
\begin{align}
    &\tilde{C}_v(\omega)-2T\tilde{R}'_v(\omega)  \nonumber \\
    &\quad=-2T \frac{ b\omega \sin(\omega \tau )} {[a+b\cos(\omega\tau)]^2+[\gamma\omega+b\sin(\omega\tau)]^2}.  \label{spectrum}
\end{align}
As shown in Fig.~\ref{fig:7}, the shape of the spectrum converges to a limiting form when $\epsilon_0\rightarrow0$ near the diffusive criticality. Near $\omega=0$, the spectrum exhibits a steep rise (or drop) with an asymptotically divergent slope for $a>0$ (or $a<0$), as seen in Figs.~\ref{fig:7b} and~\ref{fig:7c}. Expanding Eq.~\eqref{spectrum} for $\omega\ll1/\tau$ to first order yields
\begin{align}
\tilde{C}_v(\omega)-2T\tilde{R}'_v(\omega) \approx\frac{-2Tb\tau\omega^2}{\epsilon_0^2+(\gamma+b\tau)^2\omega^2}.
\end{align}
Thus, for $\omega\ll \epsilon_0/|\gamma+b\tau|$, we have $\tilde{C}_v(\omega)-2T\tilde{R}'_v(\omega)\approx -2Tb\tau\omega^2/\epsilon_0^2$. This indicates a sharp change from zero near $\omega=0$, with the spectrum reaching an extremum of approximately $-2Tb\tau/(\gamma+b\tau)^2$ when $\epsilon_0/|\gamma+b\tau|\ll \omega\ll1/\tau$. The value of $\epsilon_0$ has a limited effect on the spectrum for  $\omega\gg1/\tau$.} \red{In contrast, for parameters near the oscillatory criticality, a resonance peak emerges near $\omega=\omega_c=\sqrt{b^2-a^2}/\gamma$  in the spectrum. As shown in Fig.~\ref{fig:8}, the peak height increases as $\tau$ approaches $\tau_c=\gamma \arccos(-a/b)/(b^2-a^2)^{1/2}$ from below. To analyze this, we set $\tau=\tau_c-\epsilon_1$ with $0<\epsilon_1\ll\tau_c$ and examine the spectrum at $\omega=\omega_c+\mu$, where $|\mu|\ll \omega_c$. Analytically, expanding $\omega\tau$ to first order, the spectrum near the resonance peak at $\omega=\omega_c+\mu$ can be approximated as
\begin{align}
     &\tilde{C}_v(\omega)-2T\tilde{R}'_v(\omega) \approx -\frac{2Tb\omega_c\sin(\omega_c\tau_c)}{D_1\mu^2+D_2\mu\epsilon_1+D_3\epsilon_1^2},  \nonumber \\
     &=\frac{2T(b^2-a^2)}{\gamma[D_1(\mu-\mu_{\mathrm{max}})^2+(b^2-a^2)^2\epsilon_1^2/D_1]},
\end{align}
where
\begin{align}
    D_1&\equiv\gamma^2\omega_c^2\tau_c^2+(\gamma+a\tau_c)^2, \\
    D_2&\equiv -2\omega_c(b^2\tau_c+a\gamma), \\
    D_3&\equiv b^2\omega_c^2, \\
    \mu_{\mathrm{max}}&=-\frac{D_2\epsilon_1}{2D_1}.
\end{align}
The resonance peak has a Lorentzian line shape. Its maximum value is
\begin{align}
     &\tilde{C}_v(\omega)-2T\tilde{R}'_v(\omega)\bigg|_{\mu=\mu_{\mathrm{max}}} 
    =\frac{2TD_1}{\gamma(b^2-a^2)\epsilon_1^2}>0,\label{lorentz_height}
\end{align}
and its full width at half maximum is 
\begin{equation}\label{lorentz_width}
    \Delta\mu=\frac{2(b^2-a^2)\epsilon_1}{D_1}.
\end{equation}
These results show that as $\epsilon_1\rightarrow0$, the resonance peak becomes increasingly narrow and tall, with its integrated area diverging asymptotically. By combining Eqs.~\eqref{lorentz_height} and~\eqref{lorentz_width} and integrating the Lorentzian curve using Eq.~\eqref{hs}, we confirm that the asymptotically divergent term $2T/\epsilon_1$ in Eq.~\eqref{asymp_heat_osci} originates precisely from the integrated area of this resonance peak. 
 }
\red{}

\section{Discussion}
\label{section:5}
In this work, we have studied the heat dissipation and its properties of two classes of criticality in marginally stable linear time-delayed Langevin systems. For diffusive criticality, despite its scaled diffusion, the probability distribution of heat dissipation becomes stationary and the average approaches a constant \red{$aT(\gamma + a\tau)/\gamma(\gamma - a\tau)$} in the long-time limit. For oscillatory criticality, the distribution of heat dissipation continues to spread with time and the average asymptotically diverges as \red{$4(b^2-a^2)T\cdot t/(\gamma^2+b^2\tau_c^2-2\gamma a\tau_c)+O(1)$} due to the diffusive amplitude of oscillation. \red{Notably, the constant asymptotic value of the average dissipation rate for diffusive criticality and the linearly divergence rate for oscillatory criticality are independent of the the history trajectory.}
\red{We also analyze the asymptotic evolution of the heat dissipation rate as stable dynamics approaches these two critical boundaries, and find qualitative differences in both the convergence behavior and the associated spectral decomposition.}
\red{This analysis demonstrates that non-stationary time-delayed dynamics can exhibit qualitatively different thermodynamic behaviors, depending crucially on the underlying dynamical details.}
Our results offer a reference point for further studies on stochastic thermodynamics of more general time-delayed systems, such as those under Pyragas control and near bifurcations.

One may face both significant challenges and fruitful insights when diving into the thermodynamics of the nonlinear time-delayed regime. A promising direction is to investigate stochastic resonance phenomena on nonlinear time-delayed systems. It is natural to expect that a nontrivial resonance may manifest in their heat dissipation profiles. Furthermore, nonlinearity may suppress the amplitude diffusion of \red{Hopf} bifurcations, leading to sustained finite-amplitude oscillations of time-delayed stochastic systems. Analyzing the thermodynamic properties of such oscillations would be another path for future research. The thermodynamic properties of collective behavior \red{(e.g., synchronization or pattern formation) in many-body stochastic systems with time-delayed interaction remain an open area of research}. \red{Finally, a systematic comparison between the thermodynamics of time-delayed systems and other types of non-Markovian systems including active Brownian motion~\cite{marconi2017heat} and fractional Brownian motion~\cite{khadem2022stochastic} is also an interesting direction for future study. }

\begin{acknowledgments}
X.~W.~thanks \red{Sosuke Ito}, Naruo Ohga, Sarah A. M. Loos, and Robin A. Kopp for enlightening discussions. X.~W.~acknowledges financial support from Department of Physics, Graduate School of Science, the University of Tokyo. 
\end{acknowledgments}

\appendix
\section{Numerical methods}
We use the programming language \textit{Julia} and the package \textit{DifferentialEquations.jl}~\cite{julia_dde} and \textit{StochasticDelayDiffEq.jl}~\cite{julia_sdde} for numerical calculations and simulations (the Tsitouras 5/4 Runge-Kutta method with a free 4th order interpolant~\cite{Tsit5} is used for the numerical integration of $x_0(t)$ via Eq.~\eqref{det_eq}; the Euler-Maruyama method~\cite{EM} is used for trajectory simulations of stochastic dynamics). We draw the figures using the packages \textit{Plots.jl}~\cite{Plots} and \textit{GR Framework}~\cite{GRFramework}. 

\section{Derivation of the exponential expansion of the fundamental solution Eq.~\eqref{fun_sol_exp}}
\label{appendix:exp}
We show how to derive \red{the exponential expansion of the fundamental solution} [Eq.~\eqref{fun_sol_exp}] in detail \red{using} the method in Ref.~\cite{yi1}. Consider the linear delayed differential equation (set $T=0$ in Eq.~\eqref{par_eq}):
\begin{equation}\label{det_eq}
  \left\{
    \begin{alignedat}{2}
      &\red{\gamma}\dot{x}(t) = ax(t) + bx(t-\tau)  ,
      & (t>0) \\ 
      &x(t) = \phi_0(t).  &\hspace{-2em}(t\in [-\tau ,0])  \\ 
    \end{alignedat}
  \right.
\end{equation}
The solution of Eq.~\eqref{det_eq} is 
\begin{equation}\label{formal_sol}
  x(t) = x_0(t)\phi(0)+
  \frac{b}{\red{\gamma}}\int_{-\tau}^{0} x_0(t-s-\tau)\phi(s)ds,
\end{equation}
where the fundamental solution $x_0(t)$ corresponds to the solution of Eq.~\eqref{det_eq}
satisfying the initial conditions $x_0(t)=0$ for $t<0$ and $x_0(0)=1$~\cite{kuchler1992}.

We expand the solution of Eq.~\eqref{det_eq}  using an exponential series based on the 
roots of characteristic equation [Eq.~\eqref{char_eq}]:
\begin{equation}\label{exp_exp}
  x(t)= \sum_{k= -\infty}^{\infty}C_k e^{S_kt},
\end{equation}
where the coefficients are left to be determined according to the parameters $a, b$ and $\tau$ and the 
initial conditions $\phi(t)$. \red{Note that we have assumed $\{S_k\}$ are all simple roots, which will be justified later.} To determine the coefficients, we first compute the Laplace transform of Eq. (\ref{det_eq}):
\begin{equation}\label{Laplace}
  \red{\gamma} [sX(s)-x(0)]=aX(s)+be^{-s\tau}X(s)+b\Phi(s),
\end{equation}
where 
\begin{align}
  &X(s)\equiv\mathcal{L}[x(t)]=\int_0^\infty e^{-st}x(t)dt,  \\
  &\Phi(s)=\int_0^\tau e^{-st}\phi(t-\tau)dt, \label{Phi}
\end{align}
and the  Laplace transform of $X(t-\tau)$ is calculated as follows
\begin{align}\label{Lap0}
  &\mathcal{L} [x(t-\tau)]=\int_0^{\infty}e^{-st}x(t-\tau)dt \nonumber\\
  &=\int_0^\tau e^{-st}x(t-\tau)dt + \int_\tau^{\infty}e^{-st}x(t-\tau)dt \nonumber\\
  &=\int_0^{\tau}e^{-st}\phi(t-\tau)dt + \int_0^{\infty}e^{-s(t'+\tau)}x(t')dt' \quad (t'=t-\tau) \nonumber\\
  &=\int_0^{\tau}e^{-st}\phi(t-\tau)dt + e^{-s\tau}X(s) \nonumber\\
  &=\Phi(s)+e^{-s\tau}X(s). \nonumber
\end{align}
From Eq.~\eqref{Laplace}, we have 
\begin{equation}\label{Lap1}
  X(s) = \dfrac{\red{\gamma}x(0)+b\Phi(s)}{\red{\gamma}s-a-be^{-s\tau}}.
\end{equation}
On the other hand, the Laplace transform of the exponential expansion representation is 
\begin{align}\label{Lap2}
  \mathcal{L}[x(t)]&=\mathcal{L}\left(\sum_{k= -\infty}^{\infty}C_k e^{S_kt}\right)  \nonumber \\
 & =\sum_{k=-\infty}^{\infty}\frac{C_k}{s-S_k}. 
\end{align}
Comparing Eq. (\ref{Lap1}) with Eq. (\ref{Lap2}) and using the residue theorem, we can deduce the coefficients
\begin{align}
  C_k &= \underset{s=S_k}{\mathrm{Res}} \left(\dfrac{\red{\gamma}x(0)+b\Phi(s)}{\red{\gamma}s-a-be^{-s\tau}}\right) \nonumber \\
  &= \dfrac{\red{\gamma}x(0)+b\Phi(S_k)}{\red{\gamma}+b\tau e^{-S_k\tau}}.
\end{align}
Therefore, the exponential expansion of the solution can be written as 
\begin{equation}\label{exp_sol}
  x(t)= \sum_{k=-\infty}^{\infty}\dfrac{\red{\gamma}x(0)+b\Phi(S_k)}{\red{\gamma}+b\tau e^{-S_k\tau}}e^{S_k t}.
\end{equation}
Comparing Eqs.~\eqref{exp_sol} and~\eqref{Phi} with Eq.~\eqref{formal_sol}, we obtain the exponential expansion of the fundamental solution:
\begin{equation}
    x_0(t)=\sum_{k=-\infty}^{\infty} \dfrac{\red{\gamma}}{\red{\gamma}+b\tau e^{-S_k\tau}}e^{S_kt}.
\end{equation}

\red{We now show that for both diffusive and oscillatory criticality, the denominator of the coefficient $\gamma/(\gamma+b\tau e^{-S_k\tau})$ is nonzero, i.e. all roots of characteristic equation Eq.~\eqref{char_eq} are simple.
If a root were not simple, it must satisfy both:
\begin{align}
    &\gamma S - a - b e^{-S\tau} = 0,  \\
     &\gamma + b\tau e^{-S\tau}=0,   
\end{align}
which yields
\begin{equation}\label{double_roots}
    b = -\frac{\gamma}{\tau} e^{\frac{a\tau}{\gamma} - 1}. 
\end{equation}
For the diffusive criticality, $a+b=0$ as in Eq.~\eqref{cri_con1}; thus $ a = (\gamma/\tau) e^{a\tau/\gamma - 1}$. Setting $x = a\tau/\gamma$, we have
\begin{equation}
    x=e^{x-1},
\end{equation}
which has a unique solution $x=1$, implying $a\tau=\gamma$. This contradicts the assumption $a\tau<\gamma$ in Eq.~\eqref{cri_con1}.  For oscillatory criticality ($\tau=\tau_c$), from Eqs.~\eqref{para_x0_limit22} and setting $\omega_c\tau=\eta$, we have
\begin{equation}
    a\tau_c=\gamma \eta\cot\eta, \ \ b\tau_c = -\frac{\gamma \eta}{\sin\eta}.
\end{equation}
Combining these expressions with Eq.~\eqref{double_roots}, we obtain
\begin{equation}
    e^{\eta\cot\eta-1}=\frac{\eta}{\sin\eta}.
\end{equation}
Since $\eta = \arccos(-a/b)\in(0,\pi)$ in Eq.~\eqref{cri_con2}, $\sin\eta<\eta$ while $\eta\cot\eta<1$; therefore oscillatory criticality does not admit roots that are not simple.}

\section{Proof of the boundedness of $C_0$ and $C_1$}
\label{section:boundedness}
\red{
We explicitly prove the boundedness of the constants $C_0$ and $C_1$ in Eqs.~\eqref{C_0} and~\eqref{C_1}. We consider the overdamped time-delayed  marginally stable dynamics, where the fundamental solution $x_0(t)$ satisfies
\begin{equation}\label{fun_sol_exp_appendix} 
    x_0(t)=\sum_{k=-\infty}^{\infty} \dfrac{\gamma}{\gamma+b\tau e^{-S_k \tau}}e^{S_kt}  \ \  (t>0), 
\end{equation}
and $\{S_k\} \   (k\in \mathcal{Z})$ are all roots of the characteristic equation~\cite{yi1} 
\begin{equation}\label{char_eq_appendix}
    \gamma S_k-a-be^{-S_k\tau}=0.
\end{equation}
Subtracting from $x_0(t)$ the terms corresponding to the characteristic roots with zero real parts, we define : 
\begin{align}\label{x0_s1}
    x_{0,s}(t) &= x_0(t)-\sum_{\mathrm{Re}[S_k]=0} \dfrac{\gamma}{\gamma+b\tau e^{-S_k \tau}}e^{S_kt} \nonumber \\
    &=\sum_{\mathrm{Re}[S_k]<0} \dfrac{\gamma}{\gamma+b\tau e^{-S_k \tau}}e^{S_kt}.
\end{align}
In the following, we show the absolute convergence of $x_{0,s}(t)$ for $t>0$. Convergence at $t=0$ is ensured by the initial condition.  Using Eqs.~\eqref{x0_s1} and~\eqref{char_eq_appendix}, we obtain 
\begin{equation}\label{x0_s2}
    x_{0,s}(t) = \sum_{\mathrm{Re}[S_k]<0} \dfrac{\gamma}{\gamma-a\tau+\gamma S_k}e^{S_kt}.
\end{equation}
From the characteristic equation~\eqref{char_eq_appendix},  we have
\begin{equation}
    e^{S_k\tau} = \frac{b}{\gamma S_k - a}=\frac{b}{\gamma S_k}\cdot \frac{1}{1-a/\gamma S_k}.
\end{equation}
Taking the logarithm and expanding for large $|S_k|\red{\gg|a|/\gamma}$, we obtain
\begin{equation}\label{asymp_analy1}
    S_k\tau = 2\pi i k + \ln b - \ln \gamma - \ln S_k + \frac{a}{\gamma S_k} + O(|S_k|^{-2}),  \ \  k \in \mathbb{Z}.
\end{equation}
Let $S_k = \sigma_k+i\lambda_k$. From Eq.~\eqref{asymp_analy1}, the real and imaginary parts are
\begin{align}
    &\sigma_k \tau = \ln |b| - \ln \gamma - \ln |S_k| + \operatorname{Re}\left( \frac{a}{\gamma S_k} \right) + O(|S_k|^{-2}), \label{re_asymp1}\\
    &\lambda_k \tau = 2\pi k + \arg b - \arg S_k + \operatorname{Im}\left( \frac{a}{\gamma S_k} \right) + O(|S_k|^{-2}), \label{im_asymp1}
\end{align}
respectively \red{(for a more precise asymptotic analysis, see Ref.~\cite{wright1955non})}. Here, we take $\arg b=0$ for positive $b$, $\arg b=\pi$ for negative $b$, and $\arg S\in(-\pi,\pi]$.  For large $|k|$,  Eqs.~\eqref{re_asymp1} and~\eqref{im_asymp1}  give
\begin{align}\label{lambda_asymp1}
    |\lambda_k|  &\sim \frac{2\pi |k|}{\tau}+\red{M_0}+O(|S_k|^{-1}), \\
    \sigma_k &\sim -\frac{1}{\tau}\ln|k|+\red{K_0}+O(|S_k|^{-1}),  \label{sigma_asymp1} \\
    \red{|S_k|}&\red{=|\lambda_k|\sqrt{1+\sigma_k^2/\lambda_k^2}\sim |\lambda_k|\cdot \bigg[1+O\bigg((\ln|k|/k)^2\bigg)\bigg],}
\end{align}
where $\red{M_0}\equiv \arg b-\arg S \in(-2\pi, 2\pi]$ and $K_0\equiv \red{\ln(|b|\tau/2\pi\gamma)}$. \red{Therefore, there exists a sufficient large $N_0>0$,  such that when $|k|>N_0$, the relations at below hold:
\begin{align}
    |\lambda_k|  &\geq\frac{2\pi |k|}{\tau}-|M_0|-1, \\
    \sigma_k &\leq-\frac{1}{\tau}\ln|k|+|K_0|+1,   \label{sigma_ineq} \\
    |S_k|&\geq |\lambda_k|/2.
\end{align}
}
Taking $N_1\equiv [(|M_0|+1)\tau/2\pi]+1$ and noting that $a\tau<\gamma$, we have
\begin{equation}
     \Biggl|\dfrac{\gamma}{\gamma-a\tau+\gamma S_{k+N_1}}\Biggr|<\frac{\tau}{\pi}\cdot\frac{1}{|k|}. \label{lambda_ineq}
\end{equation}
}

\red{
We split $x_{0,s}(t)$ for $t>0$ into two parts:
\begin{align}
    x_{0,s}(t) = \sum_{|k|\leq \red{N_0}+N_1,\ \mathrm{Re}[S_k]<0} \dfrac{\gamma}{\gamma-a\tau+\gamma S_k}e^{S_kt} \nonumber \\
    +\sum_{|k|> \red{N_0}+N_1,\ \mathrm{Re}[S_k]<0} \dfrac{\gamma}{\gamma-a\tau+\gamma S_k}e^{S_kt}
\end{align}
For a specified $N_0$,  the first sum contains finite terms; hence  there exists a constant $M_1$ such that
\begin{equation}\label{sum_asymp1}
    \Biggl|\sum_{|k|\leq \red{N_0}+N_1,\ \mathrm{Re}[S_k]<0} \dfrac{\gamma}{\gamma-a\tau+\gamma S_k}e^{S_kt}\Biggr|
    \leq |M_1|e^{-\beta t},
\end{equation}
where $\beta =\underset{\mathrm{Re}[S_k]<0}{\mathrm{Inf}}
\Big|\mathrm{Re}[S_k]\Big| $. For the second sum, using Eqs.~\eqref{sigma_ineq} and ~\eqref{lambda_ineq}:
\begin{align}\label{sum_asymp2}
   &\Bigg| \sum_{|k|> \red{N_0}+N_1,\ \mathrm{Re}[S_k]<0} \dfrac{\gamma}{\gamma-a\tau+\gamma S_k}e^{S_kt}\Bigg|  \nonumber \\
   &\leq \frac{\tau}{\pi}e^{(\red{|K_0|}+1)t}\sum_{|k|\geq \red{N_0}}|k|^{-(1+t/\tau)} \nonumber \\
   &\leq \frac{2\tau}{\pi}e^{(\red{|K_0|}+1)t}\int_{\red{N_0}}^\infty x^{-(1+t/\tau)}dx \nonumber  \\
   &= \frac{2\tau^2}{\pi t}e^{(\red{|K_0|}+1-\ln \red{N_0}/\tau)t}.
\end{align}
Here we have used the condition $a\tau<\gamma$. Choosing \red{$N_0$} large enough so that \red{$|K_0|+1-\ln N_0/\tau<-\beta$}, then Eqs.~\eqref{sum_asymp1} and~\eqref{sum_asymp2}  imply the existence of a constant  $M>0$ such that
\begin{equation}\label{x_0s_bound}
    |x_{0,s}(t)|\leq M(1+t^{-1})e^{-\beta t}, \ \ \ \ \ \text{for} \ t>0.
\end{equation}
 This verifies the boundedness of $x_{0,s}(t)$ for $t>0$ and shows for large times $t$,
 \begin{equation}
     x_{0,s}(t)=O(e^{-\beta t}).
 \end{equation}
 By a similar procedure, we can also show that for the fundamental solution $x_0(t)$ of the stable dynamics [Eq.~\eqref{par_eq}] under the condition Eq.~\eqref{convergence_condition_over1} or~\eqref{convergence_condition_over2}, there exists a constant  $M_2>0$ such that
\begin{equation}\label{x_0_bound1}
    |x_{0}(t)|\leq M_2(1+t^{-1})e^{-\beta t}, \ \ \ \ \ \text{for} \ t>0.
\end{equation}
For a sufficiently large time,
 \begin{equation}\label{x_0_bound2}
     x_{0}(t)=O(e^{-\beta t}).
 \end{equation}
 Now consider the constant $C_0$:
\begin{align}
     C_0&=  \frac{2T}{\gamma}\int_0^\infty \left[ x_0^2(s)-\left( \dfrac{\gamma}{\gamma-a\tau}\right)^2 \right]ds \nonumber \\
     &=\frac{2T}{\gamma}\int_0^\infty x_{0,s}(s)\Bigg[x_{0,s}(s)+ \dfrac{2\gamma}{\gamma-a\tau}\Bigg]ds. \label{C_00}
\end{align}
The convergence speed of $C_0$ is determined by that of $x_{0,s}(t)$ \red{and is also comparable to $O(e^{-\beta t})$.}  Equation~\eqref{x_0s_bound}  ensures the boundedness on any interval $(T_0,\infty]$  with $T_0>0$ . Combined with the continuity and hence boundedness of $x_0(t)$  on $(0,T_0]$, the integral in Eq.~\eqref{C_00} is indeed bounded. A similar procedure applies to $C_1$ in Eq.~\eqref{C_1}.
}

\vspace{4\baselineskip}

%

\end{document}